%% file: main.tex
\documentclass[acmsmall]{acmart}

\setcopyright{cc}
\setcctype{by}
\copyrightyear{2025}
\acmJournal{PACMNET}
\acmYear{2025} 
\acmVolume{3} 
\acmNumber{CoNEXT4}
\acmArticle{41} 
\acmMonth{12}
\acmPrice{}
\acmDOI{10.1145/3768988}

\usepackage[utf8]{inputenc}
\usepackage{graphicx}
\usepackage{subcaption}
\usepackage{array}
\usepackage{booktabs}
\usepackage{multirow}
\usepackage[disable]{todonotes} %
\usepackage{balance}
\usepackage{wrapfig}
\usepackage{colortbl}
\definecolor{ColorGreen}{rgb}{0.59,0.73,0.38}
\definecolor{ColorRed}{rgb}{0.97,0.22,0.10}
\usepackage{soul}
\usepackage[absolute,showboxes]{textpos}

\usepackage[skip=5pt]{caption}

\usetikzlibrary{arrows,arrows.meta,external,colorbrewer,decorations.markings,decorations.pathmorphing,decorations.pathreplacing,external,fit,patterns,shapes,positioning}

\usepackage{balance}
\usepackage{hyperref}

\usepackage{sistyle}


\received{June 2025}
\received[accepted]{September 2025}

\input{commands}

\begin{document}

\date{}

\title{Waiting for QUIC: Passive Measurements to Understand QUIC Deployments}

\input{text/abstract}

\input{authors}

\renewcommand{\shortauthors}{Jonas M{\"u}cke et al.}

\maketitle

\definecolor{boxgray}{rgb}{0.93,0.93,0.93}
 \textblockcolor{boxgray}
 \setlength{\TPboxrulesize}{0.7pt}
 \setlength{\TPHorizModule}{\paperwidth}
 \setlength{\TPVertModule}{\paperheight}
 \TPMargin{5pt}
 \begin{textblock}{0.8}(0.1,0.04)
   \noindent
   \footnotesize
   If you refer to this paper, please cite the peer-reviewed publication: Jonas M\"ucke, Marcin Nawrocki, Raphael Hiesgen, Patrick Sattler, Johannes Zirngibl, Georg Carle, Jan Luxemburk, Thomas C. Schmidt, Matthias W\"ahlisch. 2025. Waiting for QUIC: Passive Measurements to Understand QUIC Deployments. 
   In \emph{Proceedings of the ACM on Networking (PACMNET) 3, CoNEXT4, Article 41 (December 2025)}. \url{https://doi.org/10.1145/3768988}
\end{textblock}

\input{text/introduction}

\input{text/background}

\input{text/method}

\input{text/analysis_stacks}

\input{text/analysis_cid}

\input{text/analysis_verification}

\input{text/related_work}
\input{text/discussion}

\input{text/conclusion}

\input{text/acks}

\label{lastpage}

\bibliographystyle{ACM-Reference-Format}
\bibliography{rfcs,ids,own,bibliography}

\clearpage
\begin{appendix}
  \input{text/appendix}

\end{appendix}

\end{document}

%% file: commands.tex
\newcommand{\result}[1]{}

\definecolor{myred}{cmyk}{0, 0.7808, 0.4429, 0.1412}

\newcommand{\done}[1]{}

\newcommand{\mytodo}[1]{}
\newcommand{\mydone}[1]{}

\usepackage{pifont}
\newcommand{\cmark}{\ding{51}}%
\newcommand{\xmark}{\ding{56}}%

\usepackage{xspace}
\newcommand{\etal}{\textit{et al.}~}
\newcommand{\eg}{\textit{e.g.,}\xspace}
\newcommand{\ie}{\textit{i.e.,}~}
\newcommand{\etc}{\textit{etc.}~}
\newcommand{\cf}{\textit{cf.,}~}
\newcommand{\one}{({\em i})\xspace}
\newcommand{\two}{({\em ii})\xspace}
\newcommand{\three}{({\em iii})\xspace}

\makeatletter
\renewcommand{\paragraph}[1]{\vspace{0.03in}\noindent{\bf #1.}\hspace{0.25ex \@plus1ex \@minus.2ex}}
\newcommand{\paragraphFirst}[1]{\vspace{0.0in}\noindent{\bf #1.}\hspace{0.25ex \@plus1ex \@minus.2ex}}
\newcommand{\paragraphNoDot}[1]{\vspace{0.03in}\noindent{\bf #1}\hspace{0.25ex \@plus1ex \@minus.2ex}}
\makeatother

%% file: text/abstract.tex
\begin{abstract}
QUIC experiences a rapid adoption since its standardization in 2021, and hypergiants configure their infrastructure to optimize for QUIC performance. 
In this paper, we introduce a passive measurement method to study both the progressive rollout and individual hypergiant configurations during the last five years. 
By analyzing backscatter traffic of the UCSD network telescope, we are able to make the following observations.
First, Meta, Google, and Cloudflare configure significantly different maximal retransmission numbers and timeouts.
Second, we can identify different off-net deployments of hypergiants, using packet features, such as QUIC connection IDs, packet coalescence, and packet lengths.
Third, we observe changing hypergiant deployment configurations during our different measurement periods.
Fourth, connection IDs can allow further insights into load balancer deployments, such as the number of servers.
We bolster our results using two orthogonal measurements: passive recording of QUIC flows and active probing.
\end{abstract}

%% file: authors.tex
\author{Jonas M\"ucke}
\orcid{0000-0003-1523-2069}
\affiliation{%
  \institution{TU Dresden}
  \city{Dresden}
  \country{Germany}
}

\author{Marcin Nawrocki}
\orcid{0000-0002-6308-5502}
\affiliation{\institution{NETSCOUT}
  \city{Westford}
  \state{MA}
  \country{USA}
}

\author{Raphael Hiesgen}
\orcid{0000-0002-1676-8108}
\affiliation{%
  \institution{HAW Hamburg}
  \city{Hamburg}
  \country{Germany}  
}

\author{Patrick Sattler}
\orcid{0000-0001-9375-3113}
\affiliation{%
  \institution{Technical University of Munich}
  \city{Munich}
  \country{Germany}  
}

\author{Johannes Zirngibl}
\orcid{0000-0002-2918-016X}
\affiliation{%
  \institution{Max Planck Institute for Informatics}
  \city{Saarbr\"ucken}
  \country{Germany}  
}

\author{Georg Carle}
\orcid{0000-0002-2347-1839}
\affiliation{%
  \institution{Technical University of Munich}
  \city{Munich}
  \country{Germany}  
}

\author{Jan Luxemburk}
\orcid{0000-0003-0879-0054}
\affiliation{%
  \institution{FIT CTU \& CESNET}
  \city{Prague}
  \country{Czech Republic}  
}

\author{Thomas C. Schmidt}
\orcid{0000-0002-0956-7885}
\affiliation{%
  \institution{HAW Hamburg}
  \city{Hamburg}
  \country{Germany}  
}

\author{Matthias W\"ahlisch}
\orcid{0000-0002-3825-2807}
\affiliation{%
  \institution{TU Dresden}
  \city{Dresden}
  \country{Germany}
}

%% file: text/introduction.tex
\section{Introduction}
\label{sec:introduction}

\begin{table}[t]
\small
\setlength{\tabcolsep}{4pt}
\caption{
Information on QUIC deployments inferred from passive backscatter traffic in 2025.
}
\label{tab:cdn-behavior}
\begin{minipage}{\columnwidth}
\begin{tabular}{llcccccccc}
\toprule
  & & \multicolumn{8}{c}{Hypergiant}
  \\ 
  \cmidrule(r){3-10} 
                 & & \multicolumn{1}{l}{Akamai} & \multicolumn{1}{l}{Amazon} & \multicolumn{1}{l}{Apple} & Cloudflare            & \multicolumn{1}{l}{Fastly} & Google                & Meta                  & \multicolumn{1}{l}{Microsoft} \\ \midrule
\multicolumn{2}{l}{First backscatter visible}    & 2023      & 2022      & 2022     & 2021 & 2023      & 2021 & 2021 & 2022         \\
\midrule
\multirow{7}{*}{\rotatebox{90}{\emph{Features}}} & Coalescence             & \cmark      & \cmark      & \cmark     & \cmark & \cmark      & \cmark & \xmark & \cmark          \\
& Structured SCIDs        & \cmark      & \cmark      & \cmark      & \cmark & \cmark       & \cmark  & \cmark & \cmark         \\
& Retry observed          & \xmark      & \xmark      & \xmark     & \cmark  & \xmark      & \xmark & \xmark & \cmark         \\
& L7 load balancers       & n/a                        & n/a                        & n/a                       & n/a                   & n/a                        & n/a                   & \cmark & n/a                           \\
& SCID length             & 20 B                       & 20 B                       & 20 B                      & 20 B                  & 17 B                       & 8 B                   & 8 B                   & 14/20 B                       \\
& Initial RTO             & 1 s                        & 0.3 s                      & 1 s                       & 1 s                   & 0.2 s                      & 0.3 s                 & 0.1 s                 & 1 s                           \\
& Mean \# retransmissions & 2.1                        & 4.0                        & 2.7                       & 1.5                   & 6.0                        & 3.4                   & 7.5                   & 1.3                           \\ \bottomrule
\end{tabular}
\end{minipage}
\vspace{-1em}
\end{table}

Measurement techniques for revealing the setups of large service providers, \ie hypergiants, are a long-standing research challenge~\cite{limoj-iit-10, bctcu-ocega-18, gcmnk-syilo-21,hsvg-flihi-24}.
Detailed knowledge of hypergiant infrastructures is often considered business sensitive and may raise security concerns.
Therefore, such knowledge is frequently hidden from the public.
Nevertheless, insights from infrastructure deployments and protocol configurations of hypergiants may help improve the experience of Internet users by fine-tuning existing deployments and guiding the development of new~protocols.

QUIC is adopted by hypergiants since 2021~\cite{zbsja-io9ae-21, mtmbb-aaoqf-20, rpdh-flaqi-18}, contributes a significant share of today's Internet traffic~\cite{c-crau-25,j-wiqg-24}, and continues to spread widely with HTTP/3~\cite{RFC-9114}.
QUIC has been designed to improve performance~\cite{cllwk-itwqd-17,wrwh-ppowo-19,sppb-eqpow-21} and to maximize privacy by disguising meta-information~\cite{vbkt-tbses-18}. 

Prior research studied the deployment of QUIC using active measurements~(\eg \cite{rpdh-flaqi-18,zbsja-io9ae-21}) or the performance and interoperability of QUIC in testbeds~\cite{jzkpc-qohep-23,si-aqit-20,mhlq-ssdds-20,mnhsw-rqmpi-24}.
Results are thus limited to lab conditions or raise operational and legal concerns due to interference with real world deployments~\cite{g-sifv-22,daswh-tyz-24,g-nnson-09,pnfb-csrim-23}. 
In this paper, we track QUIC configurations of hypergiants during the last five years using Internet backscatter traffic from the UCSD network telescope, a passive data source.
Our method is non-intrusive and does not require beforehand information on deployments.
We identify individual QUIC~configurations of Akamai, Amazon, Apple, Cloudflare, Fastly, Google, Meta, and Microsoft and gain new insights into the load balancer infrastructure of Meta, summarized in \autoref{tab:cdn-behavior}.
Furthermore, we observe the rollout process of new load balancer configurations at Meta and Google.
Our findings confirm that even when faced with metadata-hiding protocols~(\eg QUIC) analyzing traffic from network telescopes offer expressive views on the protocol ecosystem.

In detail, our contributions are as follows. %
\begin{enumerate}
    \item We introduce a new method based on passive measurements to learn about hypergiant deployments and QUIC configurations, and present results from 2021-2025 (\autoref{sec:method}).

    \item We reveal local system stack configurations of QUIC clients and servers relevant to improve performance, such as the usage of packet coalescence and retransmission behaviors~(\autoref{sec:stack-confs}).
    \item We are the first to systematically analyze QUIC connection IDs, their structure and implications in real-world deployments (\autoref{sec:cid-learnings}).
    \item We make benign use of QUIC attack traffic to detect off-net servers and show how information encoded in Connection IDs can be used to fingerprint hypergiant deployments~(\autoref{sec:cid-learnings}).
    \item We quantify and track the number of layer 7 load balancers of a single hypergiant throughout our measurement period, a previously hidden property~(\autoref{sec:cid-learnings}).
    \item We validate our results with controlled scanning campaigns and passive flow captures~(\autoref{sec:passive_and_active}).

\end{enumerate}
Our method avoids active scanning and relies on unsolicited malicious QUIC~traffic captured passively (\eg responses to requests via source address spoofing).
This allows for the observation of attacks and defense strategies  (\eg QUIC Retry packets) without interacting with attackers or infrastructure operators, therefore, our method does not trigger alarms by Intrusion Detection Systems.
Our approach captures data from a large number of deployments, which are topologically and geographically distributed.
It is difficult to ethically achieve comparable coverage with active scanning.
Many operators have a clear opinion on unsolicited traffic, such as ``I would still consider an uninvited scan of my network antisocial''~\cite{d-sifv-22}.
ISPs may even forbid port scanning in their Acceptable Use Policies, such as Xfinity~(Comcast): ``Unauthorized port scanning is strictly prohibited''~\cite{xfinity-aup-21}.
In contrast to flow measurements, our method does not require access to usually protected~data.
We do not argue that active or passive flow measurements are bad but that it is worth analyzing alternative passive options to understand hypergiant deployments based on QUIC.
To validate our results we transfer the method to flow records and perform active measurements when data is statistically sparse.
We expect QUIC backscatter to persist similar to TCP~backscatter, which has been observable for more than 25 years~\cite{hnkds-sunwo-22}, and even
to increase in the future since spoofed traffic remains an ever-increasing challenge on the Internet~\cite{c-sp-22,hnbkh-adeci-24}.

The remainder of this paper is structured as follows. We describe the problem space in \autoref{sec:background}. 
\autoref{sec:method} introduces our measurement methods and the resulting data corpus.
We report on the configurations of QUIC stacks found in the wild in \autoref{sec:stack-confs} and analyze the particular use patterns of QUIC Connections IDs and their implications for off-net servers and load balancers in \autoref{sec:cid-learnings}. We confirm the validity of our method by orthogonal measurements in \autoref{sec:passive_and_active}.
We review related work in \autoref{sec:related_work}, 
 discuss our findings in \autoref{sec:discussion}, and conclude with an outlook in \autoref{sec:conclusion}.

%% file: text/background.tex
\section{\mbox{Background}}
\label{sec:background}

This section recalls background about QUIC and discusses implications of QUIC for common hypergiant deployments.

\subsection{QUIC Overview}

QUIC implements a reliable, encrypted transport based on UDP.
A key improvement is low-latency connection establishment by combining the transport and TLS handshake into a single round-trip.
This requires the inclusion of information usually exchanged in separate TCP and TLS handshakes in the first round-trip.
Since QUIC is implemented in user-space~\cite{mhlq-ssdds-20}, several implementations with different default configurations are deployed, depending on application~needs.

\paragraph{Connection setup}
\autoref{fig:quic-connection-establishment} depicts the QUIC 1-RTT handshake.
All QUIC connections start with an \textit{Initial} packet sent by a client, which includes the TLS ClientHello~\cite{RFC-9000}.
A server replies with an \textit{Initial}~(incl. the TLS ServerHello) and a \textit{Handshake} packet~(incl. certificate and encrypted extensions), either combined in a single (\emph{packet coalescence}) or split into two separate UDP datagrams, \eg to improve the initial RTT estimate~\cite{mnhsw-rqmpi-24}.
The server will resend \textit{Initial} and \textit{Handshake} packets when the retransmission timeout~(RTO) expires due to missing acknowledgments and if the maximum number of retransmissions has not been reached.

QUIC hides metadata by encrypting type specific bits, packet number length, and the packet number. 
Nevertheless, \textit{Initial} and \textit{Handshake} packets characterize the QUIC~network stacks with cleartext version information, connection IDs, retry token, and the packet length~(see \autoref{sec:information-encoding} ).

\paragraphFirst{Connection IDs}
Clients and servers establish connection identifiers (CIDs) during the handshake, to assign packets to connections.
CIDs enable multiplexing QUIC connections over the same 5-tuple and ensure that connections survive changes in addresses or ports, \eg due to client migration.~QUIC distinguishes between source and destination IDs~(SCID, DCID), seen from the sending endpoint.

Since the client cannot guess the CID a server wants to use, it uses a temporary value (\texttt{S1} in \autoref{fig:quic-connection-establishment}), which can be replaced by the server~(ID \texttt{S2}).
SCIDs and DCIDs are unencrypted.
This enables exposure to middleboxes (\eg load balancers) and enables observing CIDs in passive measurements without knowing TLS~secrets.

Similar to TCP SYN cookies~\cite{RFC-4987}, where the TCP segment number encodes network and transport layer information to defer state allocation at servers, 
CIDs can be used to encode information in packets sent by clients.
Whenever the client returns with this information, it can be used to admit connection requests or inform loadbalancing decisions~\cite{btykm-hslbd-20,c-ugrm-25}. 

\subsection{QUIC in Hypergiant Deployment}

\begin{figure}[t]
    \centering
    \begin{minipage}[t]{0.48\textwidth}
         \centering
          \input{graphics/quic-msg-flow}
          \caption{
          Connection establishment using QUIC.
        Client connection ID (\textcolor[HTML]{000000}{C1}) is consistently used during the connection establishment but the initial server ID~(\textcolor[HTML]{f57600}{S1}) can be replaced by the server with  ID~(\textcolor[HTML]{5ba300}{S2}).}
          \label{fig:quic-connection-establishment}
    \end{minipage}
    \hfill
    \begin{minipage}[t]{0.48\textwidth}
      \begin{center}
      \includegraphics[width=\textwidth]{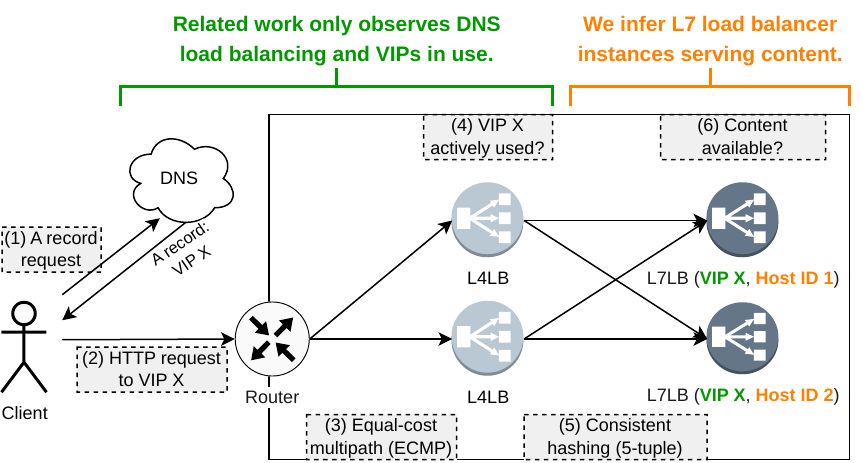}
          \caption{Illustration of a Meta \texttt{/24} load balancer frontend cluster. Our method enables L7LB quantification.} %
      \label{fig:overview_facebook_cluster}
      \end{center}
    \end{minipage}
    \vspace{-15pt}
\end{figure}

\paragraph{Common load balancer deployments}
To steer client requests to a close point of presence (PoP), many hypergiants use the DNS (see \autoref{fig:overview_facebook_cluster}) or anycast~\cite{kkhca-aicto-21, sbajr-muatm-20}.
Related virtual IP~addresses~(VIP) are assigned to multiple logical instances,  similiar to Network Address Translation (NAT) and Carrier Grade NAT~\cite{rwvab-maocn-16}.
Often VIPs belonging to a mid-size network~(\eg~/24) form a frontend cluster. 
The network belongs either to the hypergiant autonomous system (\emph{on-net deployment}) or to a third party provider (\emph{off-net deployment})~\cite{gcmnk-syilo-21,hsvg-flihi-24}.

When a client initiates an application handshake with a VIP, the handshake message is forwarded to a layer 4 load balancer (L4LB) based on equal-cost multipath.
The L4LB applies consistent hashing of the 5-tuple (\ie source and destination addresses and ports as well as transport protocol type) to tunnel the packet to a layer 7 load balancer (L7LB) via IP encapsulation.
The L7LB completes the handshake with the client and forwards the application layer request.
The number of VIPs is a poor indicator for the size of deployments~\cite{fbgkm-ttudi-21}, thus we focus on enumerating L7LBs behind VIPs.

\paragraph{QUIC-aware load balancing}
QUIC challenges common load balancing deployments for two reasons.
First, QUIC allows for client migration, but many load balancers rely on the 5-tuple \cite{eycsk-mfars-16,s-halbw-20}.
Any client that changes the source IP~address while maintaining the connection invalidates this 5-tuple, leading to a different mapping of clients to L7LB.
Instead, stateless QUIC-aware load balancers~\cite{draft-ietf-quic-load-balancers} forward traffic based on information encoded in connection~IDs.

Second, QUIC consumes and retires connection IDs (\eg during client migration).
Since new connection IDs are negotiated in encrypted packets, they must be generated by the termination point of the connection (\ie the L7LB).
Consequently, the transition from an old to a new connection ID is hidden from a QUIC-aware L4LB, preventing it from relaying packets~consistently.

To tackle this limitation, L7LBs have two options.
Either they share their active CIDs with the L4LBs, which introduces synchronization overhead between load balancers.
Or the CIDs issued by the QUIC endpoint encode the destination L7LB.
The client then uses those CIDs to contact the server, and the LBs will forward the traffic to the encoded L7LB.
The encoded information can reveal information about the server infrastructure but does not conflict with client privacy.
In \autoref{sec:cid-learnings} and \autoref{sec:passive_and_active}, we use this information to explore load balancer infrastructure.

%% file: graphics/quic-msg-flow.tex
\begin{tikzpicture}[transform shape, scale=0.85,
every node/.append style={very thick,rounded corners=0.1mm}
]
\tikzset{
  evlabel/.style = { sloped, font=\tt,above,align=center,inner sep=0.50pt },
}

\def\serverposx{6}
\def\clientposy{-0.40}
\def\msggap{-0.6}
\def\serverposy{-0.30}
\definecolor{brown}{HTML}{f57600}
\definecolor{green}{HTML}{5ba300}
\definecolor{magenta}{HTML}{000000}
\def\colorone{brown}
\def\colortwo{green}
\def\colorthree{magenta}

\node[draw,rectangle] (Client) at (0,0) {Client};

\node[draw,rectangle] (Server) at (\serverposx,0) {Server};

\draw [very thick] (Client)--(0,\clientposy+\msggap*4.5);
\draw [very thick] (Server)--++(0,\clientposy+\msggap*4.5);

\node[rectangle,draw=black,thick,fill=black!20,minimum height=0.6cm] at (\serverposx,-1.15) {};
\node[rectangle,draw=black,thick,fill=black!20,minimum height=0.6cm] at (\serverposx,\serverposy+\msggap*4.05) {};
\node[rotate=90,align=center] at (\serverposx+0.6,-1.8) {packet coalescence\\ (optional)};

\draw [->,very thick] (0,\clientposy)--node [evlabel] {Initial: DCID=\textcolor{\colorone}{S1}, SCID=\textcolor{\colorthree}{C1}}++(\serverposx,0);
\draw [->,very thick] (\serverposx,\clientposy+\msggap*0.9)--node [evlabel] {Initial: DCID=\textcolor{\colorthree}{C1}, SCID=\textcolor{\colortwo}{S2}}++(-\serverposx,0);

\draw [->,very thick] (\serverposx,\serverposy+\msggap*2*0.9)--node [evlabel] {Handshake: DCID=\textcolor{\colorthree}{C1}, SCID=\textcolor{\colortwo}{S2}}++(-\serverposx,0);

\draw [->,very thick] (0,\clientposy+\msggap*2.7)--++(\serverposx-2,0);
\node [evlabel] at (3,\clientposy+\msggap*2.6) {Init./Handshake: DCID=\textcolor{\colortwo}{S2}, SCID=\textcolor{\colorthree}{C1}};

\draw [->,very thick] (\serverposx,\serverposy+\msggap*4.1*0.9)--node [evlabel] {Initial: DCID=\textcolor{\colorthree}{C1}, SCID=\textcolor{\colortwo}{S2}}++(-\serverposx,0);
\draw [->,very thick] (\serverposx,\serverposy+\msggap*4.9 *0.9)--node [evlabel] {Handshake: DCID=\textcolor{\colorthree}{C1}, SCID=\textcolor{\colortwo}{S2}}++(-\serverposx,0);

\draw [decorate,decoration={brace,mirror,amplitude=5pt}, line width=1pt](-.1,\serverposy+\msggap*1)-- (-.1,\serverposy+\msggap*4.1*0.9) node [rotate=90, anchor=south, pos=.5, yshift=.2cm, xshift=0cm](){retransmit timeout};

\end{tikzpicture}

%% file: text/method.tex
\section{Measurement Method and Setup}
\label{sec:method}

We analyze five datasets of QUIC~Internet background radiation (IBR) from the UCSD IPv4 network telescope, each covers one month of traffic from 2021 to 2025.
QUIC flow records from a European National Research and Education Network~(NREN) and active measurements allow us to verify our observations. 
We consider two perspectives, \one QUIC stack configurations and \two larger infrastructure deployments of distributed QUIC~servers.

\vspace{-0.1em}
\subsection{Method}

\paragraphFirst{Basic idea}
We leverage data from a network telescope as passive data source.
Network telescopes capture scans (\eg by QUIC clients) and backscatter traffic (\ie replies from servers to spoofed addresses of the telescope).
We extract information available in QUIC long header packets (\ie  source and destination IP addresses and ports, the QUIC DCID and SCID, packet length(s), packet type(s)), and the reception time at the telescope.
Since its early days, QUIC shows significant IBR~\cite{nhsw-qqqrs-21}.

We mark all telescope packets with source port \texttt{UDP/443} as QUIC backscatter and all packets with destination port \texttt{UDP/443} as QUIC~requests (\ie scans).
We remove false positives based on the packet payload using the Wireshark QUIC and Google QUIC dissectors, as proposed in prior work~\cite{nhsw-qqqrs-21}, and remove data from acknowledged scanning projects~\cite{c-as-21}.
Acknowledged scanners utilize non-existing QUIC version numbers~\cite{rpdh-flaqi-18} to trigger version negotiation behavior. %
Removing these scanners prevents bias in our QUIC version~analysis.
This leaves us with less popular (\ie undocumented or unknown) and malicious scanners (\eg bots).

\paragraph{QUIC stack configurations}
We identify hypergiant on-net deployments by mapping source IP~addresses to autonomous systems (ASes) and related Regional Internet Registry~(RIR) data.
Grouping multiple QUIC packets into QUIC connections, allows inference of QUIC stack configurations, \ie retransmissions, enabled QUIC features, attack mitigations and characteristic packet features.

\paragraph{Detection of off-net server deployments}
Zirngibl \etal~\cite{zbsja-io9ae-21} revealed a fairly homogeneous infrastructure deployment of CDNs across networks. 
We use this property for identifying patterns of QUIC~traffic of on-net deployments and compare with traffic coming from non-hypergiant~ASes. 
We group similar observations into sets of off-net candidates. 
QUIC traffic features change within our observation period. 
We indicate this and adapt our off-net detection algorithms. 
For verification, we compare subject alternative names in TLS certificates of on-net deployments with those of off-net deployments. 
For this purpose we establish QUIC connections with the off-net candidates.

\paragraph{Identifying load balancers}
We identify specific load balancer instances by using information encoded in the QUIC connection ID.
Hypergiants structure this ID to include a unique host ID, which represents the actual L7LB.
We treat the connection ID structure of Meta as ground truth because it is exposed in the Meta QUIC implementation~\cite{i-m-ef-19}.

\paragraph{Verification using passive flow data}
We compare our observations in IBR with QUIC flow records, randomly sampled~(sampling rate 1:99) from inter domain links of the European NREN Czech Education and Scientific NETwork (CESNET). 
The flow records are bidirectional (\ie client requests and server responses), while IBR is always~unidirectional.

We remove incomplete and invalid flows~(\ie no TLS ClientHello parsed or missing QUIC transport parameters) exported by ipfixprobe~\cite{c-ifeds-19}.
To protect user privacy, IP addresses are exempt from this dataset. 
Nevertheless, the dataset contains the AS number, and a /24 prefix identifier of the contacted QUIC server. 
QUIC allows clients to migrate to a new address, but mandates new connection IDs in this case. 
Neither a flow exporter nor post-processing can correctly link multiple 5-tuples to the same connection since new CIDs are exchanged in encrypted packets.
Thus, a QUIC flow represents a lower bound in terms of transferred data sizes and packets. 

Each flow contains the aggregated information extracted from long header packets. 
Additionally, for the first 30 packets, QUIC packet type(s), UDP packet length, inter-arrival time, and the packet direction (\ie to server or to client) is collected.

\paragraph{Verification using active measurements}
When verifying data, we use data from third-party scans~\cite{zbsja-io9ae-21} and self-conducted active probings. 
The third-party measurement campaign combines stateful and stateless scans of the complete IPv4 space to reveal QUIC server deployments, the supported QUIC~versions, and CDN off-net deployments.
We utilize this data to identify off-net deployments where passive data is sparse.
The method of Zirngibl~\etal~\cite{zbsja-io9ae-21} of analyzing off-net deployments using QUIC transport parameters and HTTP headers remains valid to this date.
However, we find that the values of the transport parameters change within our observation period and adapt the detection accordingly.
Since 2023, we verify whether the QUIC implementations match between on- and off-net deployments of hypergiants~\cite{zgssc-qhfqd-24}.
It is noteworthy that this data quantifies VIP addresses, not L7LB instances.

We conduct active measurements to enumerate the number of layer 7 load balancers at Meta.
We connect multiple times to known QUIC servers (represented by a VIP, detected by our telescope or Zirngibl~\etal \cite{zbsja-io9ae-21}) while varying the source port of our outgoing packets.
On established connections, we log the connection IDs~(representing L7LBs).
To reduce potential conflicts of active measurements, we limit the amount of active scanning, which might lead to time delays between the (passive) observation period and active scanning.
We perform probing from a single origin within a university network.
We carefully monitor our probe for any signs of blocklisting.

\begin{figure*}[t]
  \begin{center}
    \includegraphics[width=\textwidth]{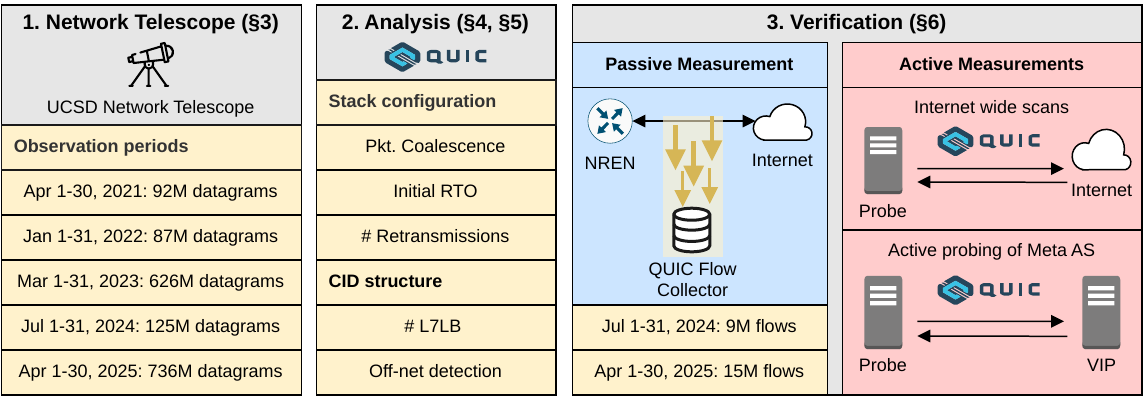}
  \end{center}
  \caption{Overview of our measurement setup and data corpus. %
  }
  \label{fig:measurement-setup}
  \vspace{-10pt}
\end{figure*}
\subsection{Data Corpus}

We utilize the UCSD Network Telescope~\cite{c-unt-12}, an IPv4 darknet operated by CAIDA within a /9 and~/10 prefix, to track QUIC IBR from 2021 to 2025.
The telescope size varies by up to~9\% in this period (\cf \autoref{sec:ucsd-changes}). 
Storage constraints limit analysis to one month of backscatter per year. 

\autoref{fig:measurement-setup} provides an overview of our measurement setup, observations, and verification.
Due to agreements with the data providers we cannot publish telescope data. 
To facilitate further research, we publish the QUIC flows sampled from CESNET and our analysis. %

QUIC backscatter traffic increased from 1M packets in 2021 to 10.8M~packets in 2023 (10$\times$) and remains stable in 2024 and 2025 with 9.1M and 10.6M packets.
This aligns with the relative development of traffic received by the network telescope.

\begin{table}[t]
\small
\caption{Number of IP addresses, and L7LBs contained in backscatter. 
We observe a strong increase in the number of VIPs, and contained L7LB-IDs. 
The number of L7LBs is determined by the number of unique host IDs per cluster in each /24 subnet.
}
\label{tab:ip-addr-host-ids}
\setlength{\tabcolsep}{1pt}
\begin{tabular*}{\linewidth}{@{\extracolsep{\fill}}lrrrrrrrrrr}
\toprule
 & \multicolumn{4}{c}{Most Backscatter Observations} & \multicolumn{6}{c}{Subsequently subsumed as \textit{Others}} \\ 
\cmidrule{2-5} \cmidrule{6-11}
 & \multicolumn{3}{c}{VIPs from Source Network [\#]} & \multicolumn{1}{c}{L7LBs [\#]} & \multicolumn{6}{c}{VIPs from Source Network [\#]} \\ 
\cmidrule{2-4} \cmidrule{5-5} \cmidrule{6-11}
Year & Cloudflare & Google & Meta & Meta & Akamai & Amazon & Apple & Fastly & Microsoft & Others \\ 
\midrule
2021 & 33 & 1,790 & 167 & 4,273 & - & 1 & - & - & - & 604 \\
2022 & 78 & 1,655 & 246 & 7,145 & 11 & 2 & 2 & - & 14 & 677 \\
2023 & 359 & 2,769 & 350 & 12,048 & 258 & 115 & 33 & 19 & 51 & 1,623 \\
2024 & 151 & 1,681 & 514 & 20,744 & 431 & 40 & 335 & 20 & 41 & 1,112 \\
2025 & 250 & 2,042 & 637 & 22,527 & 396 & 124 & 331 & 51 & 61 & 1,290 \\
\bottomrule
\end{tabular*}
\end{table}

\autoref{tab:ip-addr-host-ids} shows the number of IP addresses and L7LBs in each monthly dataset. 
In March 2023, at the largest expanse of our backscatter dataset, we observe traffic from 350~Meta, 2,769~Google, and 359~Cloudflare on-net server VIP addresses. 
This covers 7.8\%~(Meta), 1.3\%~(Google), and 0.2\%~(Cloudflare) of all VIPs that allow for QUIC connections at that time~(\cf Zirngibl \etal \cite{zbsja-io9ae-21}).
In subsequent years, the number of VIPs from Cloudflare and Google in backscatter is lower than in 2023, while the set of QUIC capable VIPs increases within that time~\cite{zbsja-io9ae-21}. 
For Meta, both the number of QUIC capable VIPs and the VIPs in backscatter increase.
In 2025, 8.7\% (637) of their QUIC capable VIPs are present in backscatter. 

Traffic from other hypergiants is present, but the number VIPs and QUIC connections~(see \autoref{sec:backscatter-prospects}) in backscatter is low, \eg 
Amazon traffic consists of a single QUIC connection in 2021.
Cloudflare, Google, and Meta contribute significantly more traffic than Apple, the next largest source, with 35,620 QUIC connections (5.9~\%) in 2025. 
Subsequently, we subsume traffic from Akamai, Amazon, Apple, Fastly, Microsoft, and all other ASes as \textit{Others}. 
The share of those providers is too small to consider them separately in aggregated statistics.

The sanitized flow records contain 449 Meta, 7,217 Google, and 5,425 Cloudflare on-net server VIP addresses in 2024, and increase to 698, 7,478, and 8,093 on-net VIPs in 2025.
Except for VIPs observed from Meta in 2024, the flow records contain more VIPs from Cloudflare, Google, and Meta than the backscatter. 
In 2025, we observe 378k flows from Cloudflare, 9.4M flows from Google, and 2.5M flows from Meta.
This is between 4$\times$ and 58$\times$ of the QUIC connections in backscatter.
We conclude, that even geographically limited vantage points can provide significant insight into hypergiant deployments. 
\autoref{sec:backscatter-prospects} includes statistics of other large content providers. 

\subsection{Limitations}
\label{sec:limitations}

\paragraphFirst{Backscatter traffic depends on attacker and target behavior}
Backscatter traffic is response traffic to packets with spoofed source addresses. 
Consequently, observations of network telescopes are limited
\one by the choice of targets of spoofers,  
\two by the mitigation efforts of the targets, and
\three by the choice of spoofed addresses and network telescope address space, \ie direct path attacks are never recorded by network telescopes.
Due to missing ground-truth, we cannot assess to which extent this impacts telescope data.
However, we observe indicators of the above limitations:
\one the majority of QUIC traffic is received by only a few subnets~\cite{mmcgm-lloln-25}, and 
\two the amount of targeted IP addresses fluctuates even at similar overall reception of QUIC traffic by the telescope.
The telescope observes significantly less VIPs compared to the active measurements~\cite{zbsja-io9ae-21}.
Despite backscatter originating from geographically distributed servers, the coverage in a region is not complete.

We are unable to dissect how mitigation efforts affect our dataset. 
Hypergiants present in one year of our dataset are contained in subsequent years. 
This only reveals that mitigation efforts, if present, do not suppress all backscatter traffic.
Nevertheless, the amount of traffic and VIP addresses from \eg Akamai and Cloudflare seems peculiarly low.
We can only speculate on attack mitigations or knowledge of telescope address space of those providers.

We are not aware of attackers polluting network telescope datasets by crafting packets resembling backscatter. 
However, telescopes would collect such forged packets. 

\paragraph{Flow records reveal global deployment configurations, and are limited to the most recent years}
Geolocating servers in flow records, we find that 87\% (2024) to 71\% (2025) of flow records communicate with servers in the United States of America. 
Flows to servers in Europe only account for 11\% (2024) to 26\% (2025).
Due to homogeneity of hypergiant QUIC deployments~\cite{zbsja-io9ae-21}, this does not limit the validity of using the flow records for verification purposes.

While the longitudinal telescope data covers 5 years, our validation data from CESNET covers only the last two years. 
Because QUIC is relatively new protocol, the flow exporter used at CESNET supports detailed QUIC flow monitoring since April 2024. 

\paragraph{QUIC exposes additional information, but analysis is limited to the handshake}
By condensing the transport and TLS handshake in a single round trip and enabling features through additional headers, QUIC exposes more information to the telescope.
Different from TCP, the first client flight includes the TLS ClientHello, which can be analyzed at the telescope. 
In response to 0-RTT packets from clients, QUIC servers can even send data with the first server flight. 
However, encryption limits observations to information from \textit{Initial} packets and cleartext packet headers~(see~\autoref{sec:information-encoding}).

%% file: text/analysis_stacks.tex
\section{QUIC Stack Configurations}
\label{sec:stack-confs}

The recent development of QUIC spread different QUIC versions, implementations, and complex protocol mechanics allow for a wide range of configurations. 
Together, this can serve as a fingerprint of the current content provider infrastructure. In this section, we analyze outstanding characteristics observable from our backscatter measurements. 

Within our measurement period, we observe both consistent and changing configurations. 
While the latter aligns with the rapid change of QUIC deployments, variations within one measurement period might hint towards heterogeneous configurations, limited amount of data, or variation in attacker behavior.
We indicate such findings but cannot dissect possible~biases. 

\begin{figure}[t]
    \begin{subfigure}[b]{.24\textwidth}
      \centering
      \includegraphics[width=\textwidth]{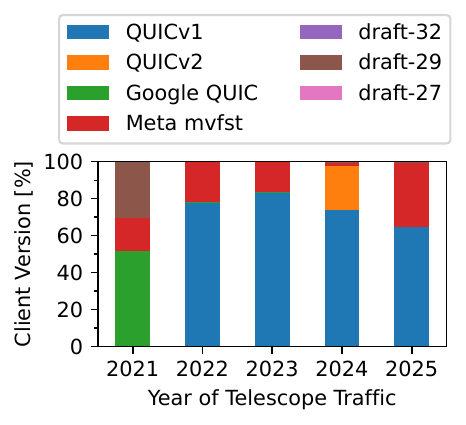}
      \caption{Clients in Telescope}
      \label{fig:quic_versions_client_telescope}
    \end{subfigure}
    \begin{subfigure}[b]{.24\textwidth}
      \centering
      \includegraphics[width=\textwidth]{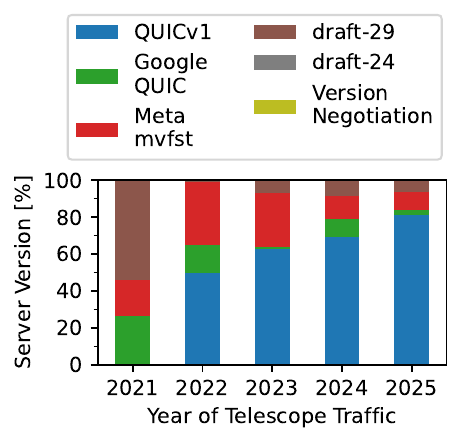}
      \caption{Servers in Telescope}\label{fig:quic_version_server_telescope}
    \end{subfigure}
    \hspace{0.01\textwidth}
    \begin{subfigure}[b]{.24\textwidth}
      \centering
      \includegraphics[width=.76\textwidth]{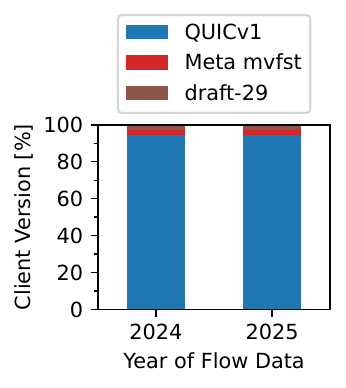}
      \caption{Clients in Flow Data}\label{fig:quic_version_client_flowdata}
    \end{subfigure}
    \begin{subfigure}[b]{.24\textwidth}
      \centering
      \includegraphics[width=0.76\textwidth]{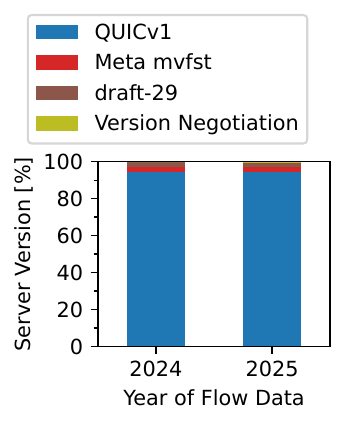}
      \caption{Servers in Flow Data}\label{fig:quic_version_server_flowdata}
    \end{subfigure}
  \caption{
    QUIC versions of clients and servers in one month of telescope traffic in 2021 to 2025, and flow records off the same months in 2024 and 2025. 
Colors included in the legend but invisible in the bars contribute $<$0.1~\% of the traffic. %
After standardization of QUIC in May 2021, QUICv1 is rapidly adopted in 2022. 
}
\label{fig:quic_versions_ibr}
\vspace{-1em}
\end{figure}

\subsection{QUIC Versions}
Whenever the version offered by a client in the \textit{Initial} is incompatible with the server, the server reacts with a 
\textit{Version Negotiation} packet indicating all supported versions. 
The client then starts a new QUIC connection with a supported version.
If the version offered by a client is compatible with the server, the server accepts the client-chosen version, or performs compatible version negotiation by transforming the \textit{Initial} to a mutually supported version. 
To allow for compatible version negotiation clients indicate supported versions in a QUIC transport parameter~\cite{RFC-9000,RFC-9368}.

\paragraph{IBR reveals that client (\ie scans) and server implementations (backscatter) adopt QUICv1 after standardization in May 2021}
QUIC IETF Draft-29 and custom versions of Google and Meta are present to this date, but clients and servers strongly gravitate towards QUICv1 (see \autoref{fig:quic_versions_ibr}).
In contrast to active scans~\cite{pdb-oeoqi-18,rpdh-flaqi-18,zbsja-io9ae-21} with non-existing QUIC versions, backscatter traffic reveals the version that client \emph{and} server agreed upon, \ie the version that is used in a connection and not only offered.
To this extend, we learn about the versions used by attackers.
QUICv2, standardized in May 2023, and discussed since Nov. 2021~\cite{RFC-9363}, is used only in 2024 in a scanning campaign originating from a Chinese NREN.
To this date, QUICv2 is not present in backscatter traffic. 

In 2025, we observe a significant increase of Client packets using of Meta mvfst versions (34.6~\%).
Those requests target a single IP address with repeating octets in the telescope.
The requests originate from ASes in Asia, with Malaysian and Indonesian ASes contributing the largest share.
We are unable to dissect the root cause for those packets.

We observe version negotiation packets from up to 19 VIPs
with Google QUIC versions but outside any Google AS and a single VIP in the Cloudflare AS in 2024.
Since version negotiation packets do not allocate state at servers, it is plausible that attackers rely on compatible versions.

\subsection{Coalescing Packets}
\begin{table}[!t]
\small
\setlength{\tabcolsep}{2pt}
\caption{
QUIC packet types visible in backscatter traffic from 2021-2025.  
Columns sum to 100~\%.
} 
\label{tab:packet-types}
\begin{tabular*}{\linewidth}{@{\extracolsep{\fill}}lllllllllllllllllllll}
\toprule
 & \multicolumn{20}{c}{Relative number of packets from source network per year [\%]} \\ 
\cmidrule(lr){2-21}
 & \multicolumn{5}{c}{Cloudflare} & \multicolumn{5}{c}{Google} & \multicolumn{5}{c}{Meta} & \multicolumn{5}{c}{Others} \\ 
\cmidrule(lr){2-6} \cmidrule(lr){7-11} \cmidrule(lr){12-16} \cmidrule(lr){17-21}
QUIC Packet Type & '21 & '22 & '23 & '24 & '25 & '21 & '22 & '23 & '24 & '25 & '21 & '22 & '23 & '24 & '25 & '21 & '22 & '23 & '24 & '25 \\ 
\midrule
\quad Initial & 42 & 56 & 54 & 49 & 45 & 34 & 23 & 7 & 9 & 35 & 65 & 48 & 47 & 43 & 47 & 69 & 46 & 33 & 36 & 43 \\
\quad Handshake & 28 & 41 & 43 & 42 & 44 & 21 & 24 & 26 & 34 & 33 & 35 & 52 & 53 & 57 & 53 & 29 & 43 & 41 & 40 & 41 \\
\quad 0-RTT & - & - & - & - & - & 2 & <1 & <1 & - & - & - & - & - & - & - & 1 & <1 & <1 & <1 & - \\
\quad Retry & - & - & - & - & 3 & - & - & - & - & - & - & - & - & - & - & <1 & <1 & <1 & <1 & <1 \\
\quad Version Negotiation & - & - & - & <1 & - & - & - & - & - & - & - & - & - & - & - & - & 3 & <1 & 3 & 1 \\
\midrule
\textit{Colaesced Packets} \\
\quad Initial+Initial & 10 & - & - & - & - & - & - & - & - & - & - & - & - & - & - & <1 & - & <1 & - & - \\
\quad Initial+Handshake & 10 & 3 & 3 & 8 & 9 & 44 & 53 & 67 & 57 & 32 & - & - & - & - & - & 1 & 9 & 26 & 20 & 14 \\
\quad Handshake+Handshake & 10 & - & - & - & - & - & - & - & - & - & - & - & - & - & - & <1 & - & - & - & - \\
\bottomrule
\end{tabular*}
\vspace{-1em}
\end{table}

QUIC allows to coalesce multiple QUIC packets into one UDP datagram.
All hypergiants except Meta use this feature (see \autoref{tab:packet-types}).
For Microsoft, we observe it since 2024 (not~shown).
Cloudflare shows a significantly lower share of coalesced packets than Google and Fastly. 
This originates from their deployment model.
The load balancer responds with a coalesced packet if it is in hold of the certificate, otherwise the acknowledgment of the client \textit{Initial} is send in a separate packet, thereby trading precise RTT-estimates for overhead \cite{nthms-ibtcq-22,mnhsw-rqmpi-24}.
We observe similarly low shares of coalesced packets from Akamai, Apple, and Microsoft (not shown).

\subsection{Packet Lengths}
\begin{figure}
  \vspace{-10pt}
  \includegraphics[width=\textwidth]{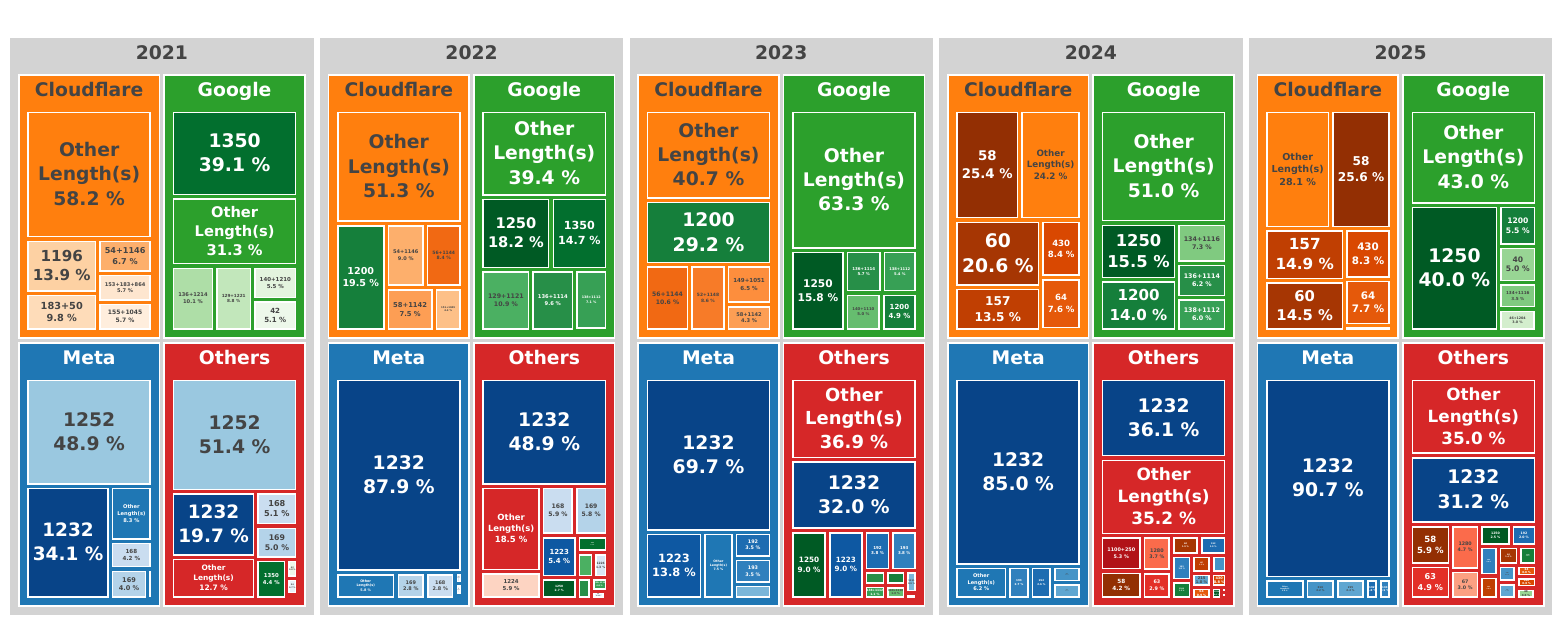}
      \caption{
        QUIC packet lengths [B] per hypergiant in 2021-2025. 
        The size of each rectangle and second line represent the proportion of the packet length indicated in the first line.
      Colors of hypergiant packet lengths in non-hypergiant ASes (\textit{Others}) point to off-net deployments. 
    }
  \label{fig:CDN-packet-length-observation}
  \vspace{-1em}
\end{figure}

\paragraphFirst{Hypergiants use distinct packet length formations}
\autoref{fig:CDN-packet-length-observation} shows the five most frequently received QUIC packet lengths in each year and combinations due to packet coalescence. 
Major providers (indicated by colors) show distinct distributions of packet length formations, which indicates this as a characterizing feature, \ie 90.7\% of packets from Meta in 2025 are 1232 B long. 
Cloudflare, Google and Meta use identical packet lengths in multiple years but with varying persistence \ie Google uses 1250~B QUIC packets since 2022, Meta uses 1232~B since 2021, and Cloudflare uses the same packet lengths only in 2024 and 2025.
We find that the Meta QUIC implementation by default uses 1232~B for sending QUIC packets~\cite{i-m-ef-19-2}.
In 2024 and 2025, Cloudflare QUIC packet lengths are much smaller compared to other hypergiants. 
This may originate from instant ACKs~\cite{mnhsw-rqmpi-24}.
Cloudflare pads those packets on the UDP level and not with QUIC padding frames.\looseness=-1 

Changes of the most frequently used packet length(s) during our measurement period, have been linked to certificate replacement~\cite{lhc-etcqc-23}, may point to changing deployments or different libraries used by clients interacting with specific services (\eg YouTube vs Instagram application).

Although the share of length(s) of \textit{Others} increased since 2023, large shares of traffic from \textit{Others} use the same packet length as Google and Meta.
The increase originates from the number of VIPs, and QUIC connection of other hypergiants are significantly increasing (see \autoref{sec:backscatter-prospects}). 
A significant share of those packet length(s) are contributed by Apple, Fastly, and Akamai in 2025. 

\subsection{Retransmission Behavior}
\begin{figure*}[t]
      \includegraphics[width=\textwidth]{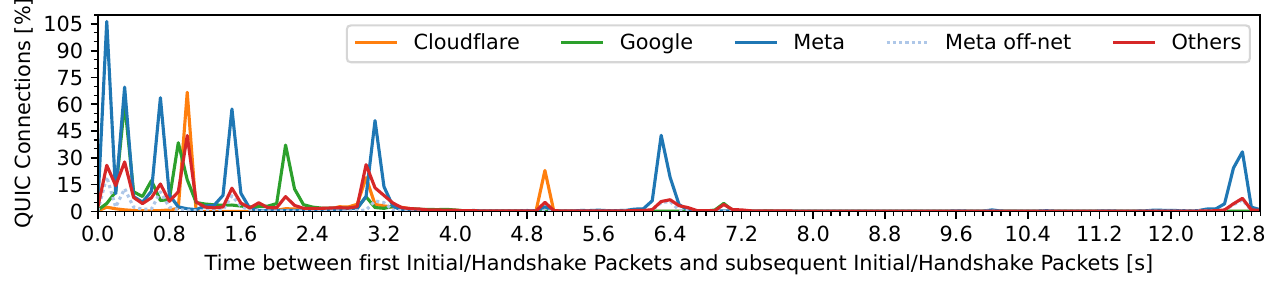}
          \caption{Retransmission configurations of QUIC servers, visible in backscatter when a server replies to a spoofed source IP~address of the telescope.
            The retransmission configurations of Google and Meta~servers are consistent throughout our observation period (shown here for 2025).
            All hypergiants use exponential backoff.
        } 
      \label{fig:retransmission-google-facebook-15s}
      \vspace{-1em}
 \end{figure*}

 A server resends both \textit{Initial} and \textit{Handshake} packets if no acknowledgment is received within the retransmission timeout~(RTO).
Unlike TCP, QUIC does not reuse packet numbers (sequence numbers in TCP), which challenges detection of retransmissions in backscatter. 
We detect a retransmission if an \textit{Initial} is received after a \textit{Handshake}.
\autoref{fig:retransmission-google-facebook-15s} depicts the time gaps between the first received packet and subsequently detected retransmissions within the same connection---resent by QUIC~servers in reply to spoofed traffic.
Peaks indicate common configurations of when and how frequently these messages were re-sent.
At 0.1~ms, we observe more retransmissions than QUIC connections from Meta. 
We find that these connections, do two resends at~\textasciitilde0.1~ms.
\begin{wrapfigure}{R}{0.5\textwidth}
  \vspace{-10pt}
  \begin{center}
  \includegraphics[width=0.5\columnwidth]{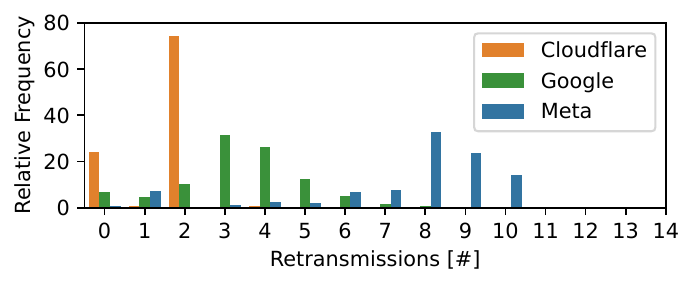}
	  \caption{Number of retransmissions
	  in 2025. 
We observe significantly more retransmission from Meta.}
  \label{fig:retransmissions-facebook-google}
  \end{center}
  \vspace{-10pt}
\end{wrapfigure}

The first retransmission happens after 0.1 s in 94\% (Meta), 0.3~s in 51\% (Google), and 1.0~s in 67\% (Cloudflare) of QUIC connections in 2025.
Except for Cloudflare (0.3~s in 31\% of the connection in 2022) this observation is stable throughout our entire measurement period, but the share of the most frequent retransmission varies. 
Possible reasons for varying configurations per hypergiant range from more heterogeneous sets of services~(Google), the low amount of data we obtained in a given year (Cloudflare), or variations in attacker behavior.
We observe exponential backoff from Cloudflare, Meta, and Google servers.
The maximum number of resends differs between hypergiants (see \autoref{fig:retransmissions-facebook-google}), showing that servers require different amounts of resources to keep connection states.
Overall, we detect the shortest resend timeouts and most retransmissions from Meta.
This indicates that Meta reacts faster to packet loss and expects shorter delays between clients and servers than Google and Cloudflare.
In comparison, Google and Cloudflare reserve less resources to cope with faulty connections.
This leads to a reduced vulnerability by QUIC flood attacks that build-up~state.

The high number of retransmissions from Meta surprises, since previous work reports amplification for Meta reduced in October 2022 \cite{nthms-ibtcq-22}. 
To validate the backscatter observation we send the same \textit{Initial} of size 1252~B seven times.
This elicits ten retransmissions at intervals matching the telescope observations.
Sending more \textit{Initials} does not increase the number of retransmissions, while sending fewer reduces the retransmissions. 
We conclude that timing of server retransmissions is not steered by clients.
Nevertheless, clients add to the byte-budget of the server with each resend.

\subsection{Denial of Service Mitigation with Retry Packets}
\paragraphFirst{DoS mitigation with \textit{Retry} packets is rarely used} 
QUIC \emph{Retry} packets enable QUIC servers to verify the client address by mandating the client to reconnect with a Retry Token given by the server. 
Those \emph{Retries} can be effective against QUIC floods but add an RTT~\cite{nhsw-qqqrs-21}.
We observe this defense strategy rarely deployed (see \autoref{tab:packet-types}).
For example, Cloudflare started in 2025, where 3\% of QUIC packets from Cloudflare are \emph{Retries}).
This indicates that deployments favor low latency connection setups over DoS mitigation.

%% file: text/analysis_cid.tex
\section{What can Connection IDs tell?}
\label{sec:cid-learnings}

To reduce tracking vectors, QUIC connection IDs must not contain information that allows correlation of multiple CIDs to one QUIC connection \cite{RFC-9000}. 
When deliberately chosen, their unencrypted transfer allows loadbalancers to forward packets to the same L7LB even across changing 5-tuples.
In this section, we analyze the CIDs chosen by clients and servers, use their structure to detect off-net deployments, and infer the number of L7LBs behind VIPs. 

\subsection{Structure of Client CIDs}
\paragraphFirst{Client CIDs are frequently zero-length or set at random}
QUIC servers use the SCID set by the client in the first flight as the DCID in subsequent packets. 
Until 2023, 99\% of the DCIDs in long header packets in backscatter were zero length. 
In this case clients must not use the same source port and address for multiple QUIC connections.
The proportion of zero length DCIDs reduced to 71\% in 2024 and in 88\% in 2025.
This originates from Cloudflare, Apple, and Akamai backscatter, which send an 8 byte DCID in 95-100\% of QUIC long header packets.
We find that client CIDs follow the random distribution - not revealing any information.

\subsection{Structure of Server CIDs}
\begin{figure*}[t]
      \includegraphics[width=\linewidth]{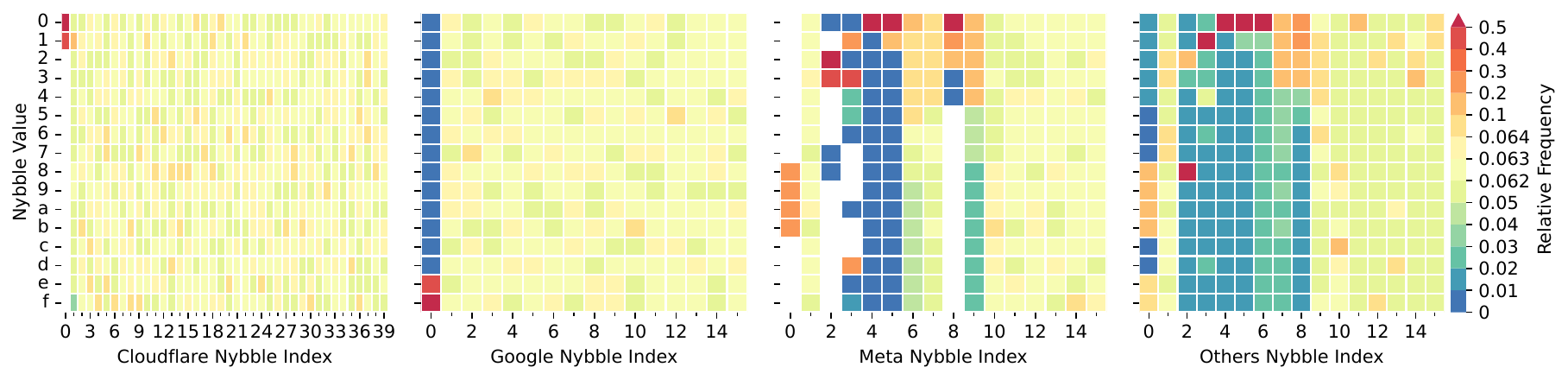}
  \caption{Relative frequencies of SCIDs values in backscatter from 2024. 
    A non-uniform distribution per column indicates information encoding (\eg Google). 
    \textit{Others} contains patterns similar to those of Meta and~Google.
  }
  \label{fig:scid-nybble-values}
  \vspace{-15pt}
\end{figure*}

To ensure high entropy despite information encoding and to prevent collisions between multiple connections, we found server CIDs between 8 and 20~bytes are preferred by hypergiants over shorter or zero-length CIDs.
SCIDs (\cf S2 in \autoref{fig:quic-connection-establishment}) can leak data if hypergiants encode information in them.
Such encoding distorts the uniform distribution of specific values in the SCID.
\autoref{fig:scid-nybble-values} visualizes the frequencies of SCID nybble values as monitored in the backscatter traffic.
Frequencies that diverge from the random distribution, \ie expected value $\frac{1}{16}=0.0625$~(light-yellow and -green) show SCIDs encoding specific information.

We observe that Akamai, Amazon, Apple, Google, Fastly, Meta, Microsoft and Cloudflare repeatedly use reoccurring values at specific positions within the SCID, indicating information encoding.
From Akamai, Amazon, Apple, and Cloudflare we observe 20~B SCIDs, Fastly uses 17~B SCIDs, Google and Meta use 8~B SCIDs, and Microsoft uses 14~B and 20~B SCIDs.

\paragraph{Cloudflare SCIDs}
Since the beginning of our measurements Cloudflare uses 20~B connection IDs and a fixed first byte \texttt{0x01}.
Our large-scale active measurement data set confirms their preference for this fixed first byte and, in a prior post~\cite{j-ghswq-18}, Cloudflare acknowledged the benefits of encoding information in QUIC~IDs.
The encoded information is not described in their open source QUIC implementation \textit{quiche}~\cite{c-q-18} or by the IETF draft for Generating Routable QUIC Connection IDs \cite{draft-ietf-quic-load-balancers}, though, as the first byte would indicate a connection ID length of 1 or include random bits.

\paragraph{Google SCIDs}
In 2021 and 2022, the SCIDs from Google follow a random distribution. 
In 2023 and 2024, the loadbalancer configuration changes.
We observe SCIDs starting with \texttt{0b11} in 99\% (2023) and \texttt{0b111} in 99.9\% (2024) of long header packets.
By manually connecting to Google servers, we find that Google echoes the first 8 bytes of the DCID provided by clients and overrides the first bits since adopting information encoding.
This means that our backscatter exposes what clients send to Google, and clients should send random connection IDs (\autoref{sec:background}).
Echoing large parts of the CID weakens \textit{Initial} protection, but does not impact the setup of the encrypted connection.

The Google QUIC implementation QUICHE supports the IETF Draft for Generating Routable QUIC Connection IDs \cite{draft-ietf-quic-load-balancers,g-q-21}.
\texttt{0b11} and \texttt{0b111} indicate loadbalancers to forward according to the 4-tuple instead of information from the CID.
IETF Draft-18 \cite{draft-ietf-quic-load-balancers-18} changed from two bit encoding to three bit encoding.
Surprisingly, Google closely tracks the development of the draft, while we do not detect indications for CID based loadbalancing.

\begin{figure}[t]
    \centering
    \begin{minipage}[t]{0.48\textwidth}
        \centering
            \includegraphics[width=\textwidth]{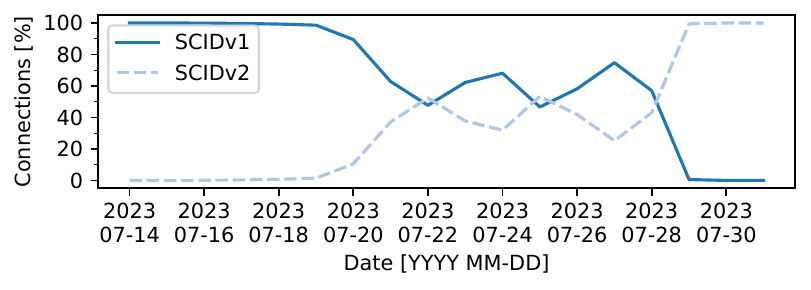}
        \caption{Encoded SCID version in Meta QUIC connections in backscatter. Between July, 19 and  July, 29 2023 Meta migrates to SCIDv2.}
        \label{fig:lp-mvfst-scid-version-usage-flows}
    \end{minipage}
    \hfill
    \begin{minipage}[t]{0.48\textwidth}
      \begin{center}
          \includegraphics[width=0.75\textwidth]{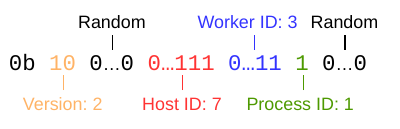}
          \caption{Meta encodes version, host, worker and process IDs into the CID.  See \autoref{sec:facebbok-connectionids} for details on all versions of CID encoding.}
      \label{fig:meta-cid-encoding}
      \end{center}
    \end{minipage}
    \vspace{-1\baselineskip}
\end{figure}

\paragraph{Meta SCIDs}
The Meta QUIC~implementation \textit{mvfst}~\cite{i-m-ef-19} allows for encoding details about hosts, workers, processes, and the version of this encoding within the SCID (see \autoref{fig:meta-cid-encoding}).
Given higher densities for some values in the first five bytes we conclude that Meta currently encodes information.

Closely tracking the reception of QUIC backscatter from Meta allows us to observe the migration to a new loadbalancer configuration, from SCID version~1 to version~2, in July 2023 (see \autoref{fig:lp-mvfst-scid-version-usage-flows}).
The migration took 10 days, and follow-up active measurements confirm the changes.

Before this migration, host IDs denominated individual L7LBs. 
Thereafter, Meta uses the same host IDs in different clusters. 
We detail our method of cluster inference in \autoref{subsec:verification_active}.
According to it, the number of L7LBs in backscatter increases from 4,273 in 2021 to 22,527 in 2025. 
Compared to active measurements, backscatter observes the largest proportion of Meta host IDs in 2023 with 29\%.

\subsection{Passive Detection of Off-net Servers}
\label{sec:facebook-offnet-classification}

\paragraphFirst{All hypergiants use information encoding in connection IDs}
Entirely passive, we are able to determine that all hypergiants observed in backscatter use information encoding in CIDs.
We now use the aforementioned distinct patterns to detect off-net deployments. 
Methods from prior work, using subject alternative names from certificates~\cite{gcmnk-syilo-21} and transport parameters~\cite{zbsja-io9ae-21} cannot be transferred to backscatter, since this information is exchanged in encrypted \textit{Handshake} packets.

\paragraph{Evaluation method}
We analyze all QUIC~servers emitting backscatter that are deployed in non-hypergiant ASes.
To evaluate our approach, we need to associate off-net IP addresses with hypergiant services. 
We actively connect to candidates and inspect X.509~certificates.
We label IP addresses as off-net deployment if the subject alternative names include any of the domains \textit{facebook.com, instagram.com, fbcdn.net, whatsapp.com, whatsapp.net} for Meta and \textit{google.com, googlevideo.com, doubleclick.net, edgestatic.com, gstatic.com, blogger.com, googleapis.com, gvt$[$0-9$]$.com} for Google.
We only present data since 2022, since we collect certificate data since then.

We classify an IP address (\ie off-net candidate) operated by a given hypergiant if all SCIDs issued by that IP address conform to the passively detected pattern, \ie \texttt{0x01} for Cloudflare,
\texttt{0b01} for \textit{Google SCIDv1},
\texttt{0b111} for \textit{Google SCIDv2},
\texttt{0b01} for \textit{Meta SCIDv1},
and \texttt{0b10} for \textit{Meta SCIDv2}.

\begin{wrapfigure}{R}{0.52\textwidth}
  \vspace{-10pt}
  \captionof{table}{
$F_1$-score of SCID classifiers. 
  Low $F_1$-scores for Google in 2022 originate from not using information encoding. 
Meta off-net classifiers consider more bits than Google classifiers and achieve better scores.
} 
\label{tab:f1-score}
\begin{tabular}{@{\extracolsep{\fill}}lcccc}
\toprule
 & \multicolumn{4}{c}{$F_1$-score} \\ 
\cmidrule(lr){2-5}
Classifier & 2022 & 2023 & 2024 & 2025 \\ 
\midrule\addlinespace[2.5pt]
Meta Off-net SCIDv1 & 0.98 & 0.98 & - & - \\
Meta Off-net SCIDv2 & - & - & 0.98 & 0.99 \\
Google SCIDv1 & 0.17 & 0.89 & 0.79 & 0.77 \\
Google SCIDv2 & 0.12 & 0.38 & 0.8 & 0.78 \\
\bottomrule
\end{tabular}
\end{wrapfigure}
\paragraph{Evaluation results}
Subsequently, we show results from 2023 because backscatter contains the most off-net candidates in this year.
For Google, this predicts 556 candidates.
Conservatively, we mark 100 of those as false positives as they do not allow for QUIC connections, \ie 456 are true positives.
1516 are predicted negative---only 7 are false negatives (TPR 0.98, FPR 0.06).
For Meta, we identify 727~candidates. 
651 of those indeed belong to Meta. 
22 do not allow QUIC connections, we consider them as false positives and 54 are falsely predicted as Meta off-net deployments.
No false negatives are predicted (TPR 1.0, FPR 0.05).
We find six off-net candidates for Cloudflare, but none allow for QUIC connections, preventing us from fetching certificates to determine ground truth.

\paragraph{Improving SCID-based detection}
We find that Meta off-net servers use low numbers for host IDs in 2023.
Active probing of 45k Meta off-net VIPs found by Zirngibl \etal \cite{zbsja-io9ae-21} confirms this.
Consequently, requiring the first 9~bits of the host ID to be zero confirms the 651 off-net deployments and reduces the false positive rate (TPR 1.0, FPR 0.02).

Besides SCID structure, we perform off-net detection based on retransmission intervals, packet lengths, and coalescence, as well as combinations of them. 
We detail those in \autoref{sec:classifiers}.

\paragraph{SCID classifiers are effective in all years with structured connection IDs}
The applicability of our method is not limited to a single year of observations. 
Hence, we show the $F_1$-score to evaluate predictive performance in \autoref{tab:f1-score}.
The $F_1$-score combines precision and recall into a single value. 
The maximum attainable value is 1, $F_1$-scores larger than 0.9 are generally considered very good.
We achieve very good performance using the Meta off-net classifiers, while performance of the Google SCID classifiers is worse due to using fewer bits.
VIPs with few connections are misclassified as Google if their SCID matches the Google SCID structure by chance. 
This increases the number of false positives and negatively impacts the precision.
Notably, very low F1 scores for Google in 2022 originate from not using information encoding in this year.

%% file: text/analysis_verification.tex
\section{Verification}
In this section we \one assess the correctness of observations in backscatter with QUIC flow records from CESNET, \two enhance the information extracted from backscatter to assess completeness of backscatter observations and gain detailed insight into L7LB deployments at Meta.

\label{sec:passive_and_active}
\subsection{Verification with Flow Records}

\paragraphFirst{QUIC versions}
Each flow stores the version indicated by the client in the Client \textit{Initial}, and the  version the client and server agreed on. 
The majority of clients and servers use QUICv1, while a small share of the connections is performed with a custom version of Meta. 
A QUIC draft version and version negotiations messages are also present, but without a significant share (see \autoref{fig:quic_versions_ibr}). 
Flow records confirm backscatter observations.

\begin{wrapfigure}{R}{.5\textwidth}
  \vspace{-10pt}
  \centering
  \includegraphics[width=0.5\textwidth]{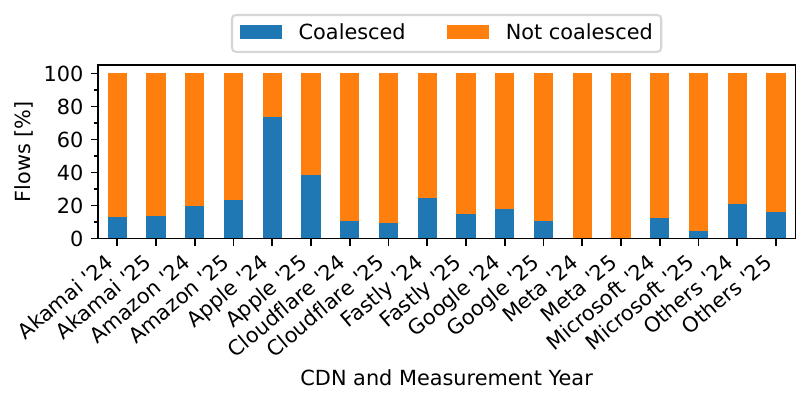}
  \caption{Share of packet coalescence in server packets. Meta servers do not use packet coalescence, while packet coalescence is used by all other CDNs.}
  \label{fig:packet-coalescence}
  \vspace{-10pt}
\end{wrapfigure}
\paragraph{Coalescing packets}
Flows store flags for QUIC packet types in the first 30 UDP datagrams. 
This reveals coalescence of different packet types, but coalescence of the same packet type~(\eg two Initial packets) remains hidden.
At present, coalescence of different packet types is the most common (\cf \autoref{tab:packet-types}).

\autoref{fig:packet-coalescence} shows 10\% (Cloudflare) and 18-10\% (Google) of datagrams from servers contain coalesced packets. 
This amounts to less coalesced packets in flow records compared to backscatter for Google (\ie up to 39\% less) and similar proportions for Cloudflare.
Lower observation of packet coalescence is expected for flow records because reactive clients do not necessitate the server to resend information from the QUIC handshake, which is composed of Initial and Handshake packets.

\paragraph{Packet lengths}
Flows collect the UDP payload length for the first 30 packets.
The most common size of server packets is $\ge$~1200~B from Google and Meta.
The sizes match with the sizes observed in backscatter, although other packet length are also present \eg~1357~B.
For Cloudflare the flow record and telescope observations deviate.
The most common packet length from Cloudflare servers is 1200~B, followed by small packets (<52B).
Small packets are similar but not identical to packet sizes in backscatter.
1200~B is present in backscatter, but at a very low share.

\paragraph{Retransmission behavior}
Backscatter contains server reactions to unresponsive clients. 
To track such behavior in flow records, we select flows that consist of a single packet sent by a client and at least three packets sent by the server.
Those records occur when clients abandon connections after sending a request, \eg users close a web browser before the QUIC connection is established.
Retransmission in flow records match the retransmissions observed in backscatter (not shown).

\paragraph{Client CIDs}
Zero-length connection IDs are the most common CID lengths (75\% flow records), which similarly appears in backscatter traffic of those years.
16~B and 3~B connection IDs are the next most common CID lengths.
Testing browser connections to QUIC servers, we find that Chrome and Safari use 0~B SCIDs while Firefox uses 3~B SCIDs.

\paragraph{Server CIDs}
The flows reveal the same information encoding as backscatter (\cf~\autoref{fig:scid-nybble-values}), \ie confirms structured CIDs of Akamai, Amazon, Apple, Cloudflare, Google, Meta and Microsoft servers.
However, we find significant shares of other CID length in flows from Akamai, Amazon, and Microsoft. 
Only one connection ID length from Amazon shows no signs of information encoding. 

\paragraph{Meta CIDs}
Counting unique host IDs per /24 prefix, we detect 10,038 L7LBs in 2024 and 25,534 in 2025---between 3\% and 7\% of the total number or L7LBs in active probing of Meta on-net servers (see \autoref{subsec:verification_active}).
The SNI in each flow record, allows us to use the same domains to find Meta off-net servers as described in \autoref{sec:facebook-offnet-classification}. 
This confirms 1,003 off-net VIPs.
Even our geographically limited vantage point reveals a significant number of L7LBs and off-net deployments.

\paragraph{Except for packet length(s) observations, flow records confirm insights from backscatter}

\subsection{Verification of Convergence: Active Measurements to Improve Statistics}
\label{subsec:verification_active}

Using active scans towards hypergiants between 2022 and 2025, we now assess the robustness of backscatter observations and extend our understanding of hypergiant QUIC deployments. 
While we want to infer how much backscatter is needed to gain statistical convergence, we concentrate on Meta since only Meta confirms their encoding of L7LB host IDs in their SCIDs (ground truth). 
To this end, we scan all QUIC VIPs active at the time (up to 7,355 VIPs in 2025) in the Meta AS~32934.
For each VIP, we complete 20k~handshakes while successively decreasing the client port.

\paragraph{Meta clusters 2022-2023}
We group VIPs into clusters, when the same host ID is used by multiple VIPs in the same /24 prefix.
We confirm the derived structure of clusters using reverse DNS, the IATA airport code encoded in DNS PTR records of all cluster VIPs is identical, \ie clusters are limited to a single /24.
Using this method reveals a set of \textit{large clusters}~($\ge 20$ VIPs) and \textit{small clusters}~($\le$ 10 VIPs).
In 2022, we find 114 clusters composed of 22 VIPs and 1 cluster composed of 20 and 21 VIPs in 2022 (total: 116).
In 2023, the number of \textit{large clusters} and VIPs within each cluster increases. 
We now find 120 clusters with 27 VIPs and 1 cluster with 22 and 28 VIPs in 2023 (total: 122). 
Within a single cluster we observe up to 461 different host IDs, \ie L7LBs.
Domains hosted by the \textit{large clusters} cover the full range of Meta services. 

Similar to large clusters, the number of \textit{small} clusters and size of clusters increases from 43 clusters with 8 VIPs, 2 clusters with 7 VIPs and 1 clusters with 6 VIPs in 2022 (total: 46) to 55 clusters with 10 VIPs, 5 clusters with 9 VIPs, and 19 clusters with 1 VIP in 2023 (total: 79).
The clusters with 1 VIP have up to 225 L7LBs, while all other \textit{small clusters} are composed of up to 16 host IDs.
When collecting certificates from the VIPs, we found all of the cluster with more than 1 VIP serve \textit{fbcdn.net} except for a single VIP in each cluster to serve \textit{whatsapp.net}. 

\paragraph{After rollout of a new loadbalancer configuration host IDs are repeated in multiple clusters}
In 2022 and beginning 2023, host IDs were often globally unique. 
After the rollout of \textit{Meta SCIDv2} in July 2023, we observe same host IDs now in different clusters.
We confirm this by active measurements in the beginning and end of July 2023. 
In early July, 3\% of the VIPs used \textit{Meta SCIDv2} and 41,016 host IDs per /24 prefix (31,733 unique host IDs) are detected. 
On 31 July, 95\% of all Meta on-net VIPs use \textit{Meta SCIDv2} and the number of L7LBs is 41,277 (4,193 unique host IDs).
The migration did not change the number of L7LBs, but host IDs are now repeated.

\paragraph{Meta clusters 2024-2025}
In 2024, we see significantly higher numbers of smaller clusters, \ie  1151 (2024), and 1532 (2025).
No clusters with more than 10 VIPs exists. 
We observe 65 clusters with 10 VIPs in 2024 and 82 in 2025;  120 (2024) and 116 (2025) with 9 VIPs.
Furthermore, the share of clusters with 2-8 VIPs contributes 428 clusters in 2024 and 788 in 2025.
There also are clusters with a single VIP (538 in 2024 and 546 in 2025).
The largest cluster is composed of 466 host IDs in 2024 and 1360 in 2025.
Collected certificates do not provide distinct sets of domains, which would explain the new structure. 
PTR records reveal, clusters with 1 VIP tend to point to WhatsApp, while 9-10 VIPs point to the whole range of Meta services. 
We also see significantly more PTR records containing \textit{edge}, which may indicate an increased number of edge locations. 

\begin{figure*}[]
  \includegraphics[width=1\textwidth]{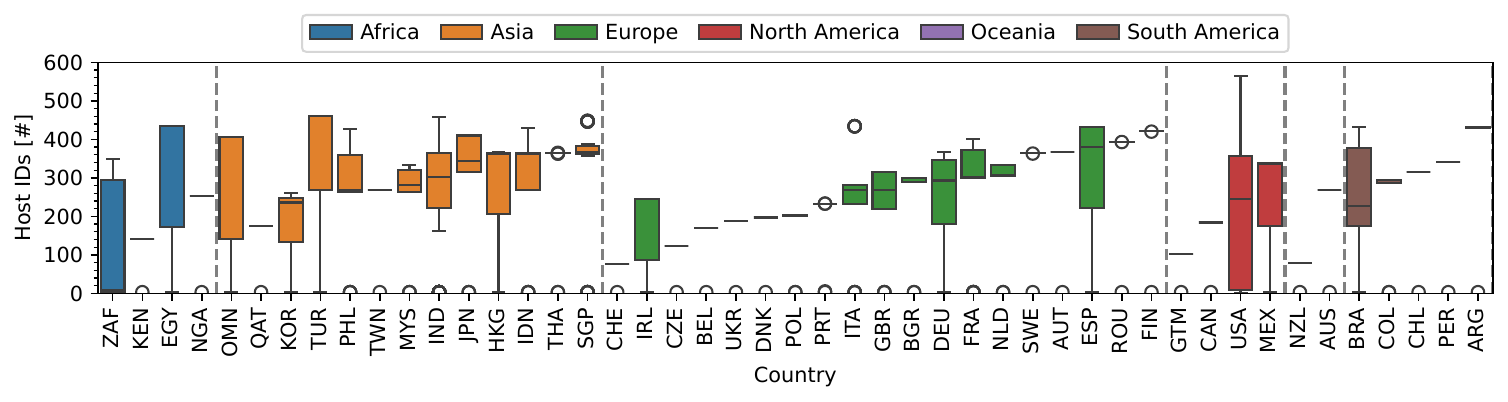}
  \caption{Meta L7LBs aggregated by country in 2025. PoPs in the Asian region utilize more L7LBs~(\# host IDs).}
  \label{fig:boxplot-countries-host-count-colored-by-continent}
  \vspace{-2em}
\end{figure*}

\begin{wrapfigure}{R}{.5\textwidth}
  \vspace{-10pt}
  \centering
    \includegraphics[width=0.5\columnwidth]{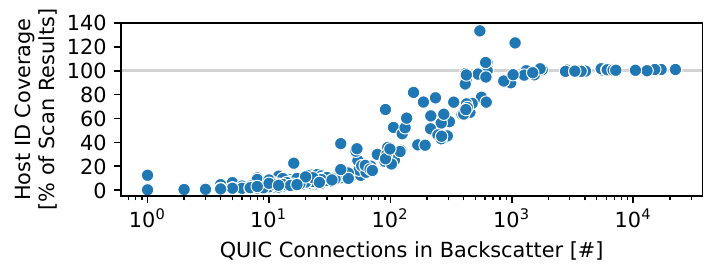}
  \caption{Number of host IDs in Meta clusters visible in backscatter relative to hosts IDs in active measurements. 
    With increasing number of QUIC connections a larger percentage of L7LBs is discovered.
For 21 clusters, we observe more host IDs in backscatter than during our active measurements, \ie >100\%.}
  \label{fig:backscatter-visibility}
  \vspace{-10pt}
\end{wrapfigure}
\paragraph{Meta on-net clusters are large in Asia}
\autoref{fig:boxplot-countries-host-count-colored-by-continent} depicts the number of L7LBs per cluster in different countries.
Until migration to \textit{Meta SCIDv2}, \textit{small clusters} were only located in North America~(46) and Africa~(15).
We find that the median number of L7LBs in \textit{large} clusters is significantly higher in Asia than on any other continent~(414 \textit{vs.} 344.5 in Europe and 353.5 in North America in 2023).
In 2024, the clusters change significantly. 
In 2025, 580 clusters are located in Asia---significantly more than North America~(395) and Europe~(347).
The median number of L7LBs per cluster in Asia~(317) continues to be larger than in Europe~(269), 
North America~(184) and South America~(295).

We conjecture three reasons for this:
\one the number of available peering points in Asia is limited,~
\two political instabilities in specific regions and regulations limit additional data centers, and~
\three high population and hence user densities per region.
All three reasons may guide Meta to not increase the number of PoPs but to provide more L7LB instances per PoP.

\paragraph{95\% of host IDs are discovered within 1,000 connections}
On average, we detect \textasciitilde95\% of all host IDs after 1k~handshakes per~VIP.
 Also, 99\% of the VIPs aggregate the same number of host IDs~(deviations $\leq$ 5)  within each cluster.  
We conclude that it suffices to observe backscatter or probe one VIP to determine the number of all L7LBs of one cluster.

\paragraph{Coverage of backscatter} 
We compare the number of host IDs in backscatter with our active measurements. 
\autoref{fig:backscatter-visibility} depicts that more QUIC connections lead to larger coverage of the number of host IDs.
While in 2023, 2,366 QUIC connections allowed detection of 93\% of all host IDs in a cluster, we achieve 100\% coverage with only 545 QUIC connection in 2025 following the new structure.
This is related to the on median lower number of L7LBs per cluster.
For 21 clusters, we observe more host IDs in backscatter than in our active measurements. 
This may be due to cluster reconfiguration during the observation period or delays between the passive observation and active probings, in which instances may be replaced or clusters extended.

%% file: text/related_work.tex
\section{Related Work}
\label{sec:related_work}

\paragraphFirst{QUIC Deployments}
Prior studies measure QUIC hypergiant deployment by counting VIPs through active measurements~\cite{rpdh-flaqi-18,zbsja-io9ae-21}, evaluate spread of early QUIC versions in MAWI traffic data, netflow from a European Tier-1 ISP, and a European IXP~\cite{rpdh-flaqi-18} or evaluate implementation behavior, interoperability, and performance of public test instances and sandboxed deployments \cite{mhlq-ssdds-20,si-aqit-20,jzkpc-qohep-23,zgssc-qhfqd-24,mnhsw-rqmpi-24,ts-cmiyc-21}.
In contrast to prior research, our work uses less intrusive passive measurements (QUIC backscatter traffic) and is more accurate by counting L7 loadbalancers instead of VIPs.
We extend telescope usage from quantifying protocol attacks to understand QUIC configurations in the~wild by leveraging standardized protocol mechanics and traffic solicited by malicious actors. 

\paragraph{Stack Configurations}
Related work~\cite{mmmr-pata-08,jt-aibth-07} focuses on the detection and classification of TCP anomalies not on the detection of individual stack configurations. 
They use testbeds or traffic traces from real networks but not darknet traffic provided by telescopes.
To the best of our knowledge, no analysis of stack configurations (\eg inference of retransmission timeouts and number of retransmissions) neither on QUIC nor TCP has been performed. 
We consider this future work.

\paragraph{Inferring Loadbalancers}
Measurement methods that identify loadbalancers or larger CDN infrastructure use traceroute-like tools (\eg \cite{acovf-atawp-06,vrbff-dcdoi-20}) or build on IP~addresses made available via the DNS~(\eg~\cite{amsu-wcc-11,drhb-evoco-22}).
Similarly, machine-learning techniques for traffic classification of TLS and QUIC \cite{scfc-mftih-16, ttsm-nqtcb-18, yfjr-dlazt-21, asvlb-lbctc-21, lhc-etcqc-23} and TLS fingerprinting using handshake variations \cite{hcjc-htaac-16} are used to derive services and hypergiant deployments.
The same services deployed as off-nets can thus reveal off-net footprints.
While active probing, DNS, and traffic classification can quantify the number of VIPs they do not identify L7LBs.
\cite{ktf-qalba-25} count L7LBs through successful QUIC connections until a failure occurs when reusing connection IDs. 
Since loadbalancing does not guarantee all L7LBs receive a connection before forwarding to the same L7LB again, the number of L7LBs is only a lower bound. 
We use information encoded in connection IDs to unveil the number of L7LBs and structure of frontend clusters at Meta, a previously hidden property. 

%% file: text/discussion.tex
\section{Discussion}
\label{sec:discussion}
In this work, we extracted various deployment details from meta-information encoded in QUIC~headers.
Although QUIC was designed to hide meta-information, we find that efficient loadbalancing and client migration require additional information encoding.
This does not leak details about clients but about the server infrastructure of hypergiants.

\paragraphNoDot{Why use passive measurements?}
With our analysis we show that 
\one even small amounts of backscatter reveal QUIC stack configurations of hypergiants, and \two enumeration of loadbalancer instances is possible at high fidelity, with less than 1,000 spoofed QUIC connections in backscatter. 

Any observation from passive measurements is reproducible with active measurements, but active measurements require prior knowledge on potential targets, cost additional network traffic, and might trigger intrusion detection.
Telescope measurements can help to improve active measurements as they shed light on real world QUIC traffic.
Given the variety of existing QUIC~libraries and their potential configurations, reproducing realistic behavior for active measurements can be challenging.
Insights from passive data should be used to guide future scan~campaigns.

\paragraphNoDot{Is passive data biased?}
Network telescopes capture scans and responses to spoofed traffic.
This does not provide insights into Web clients but into QUIC~servers and software of malicious actors.
First, it exposes QUIC~clients that are used for benign or malicious scans.
Second, additionally to measuring QUIC~features current server deployments offer, we measure what clients and servers agree on when they communicate.
Spoofed traffic is primarily triggered by malicious activities (\eg botnets), which allows us to indirectly derive insights into software deployed in such environments.

Our telescope does not receive traffic from all PoPs and VIPs, but we were able to show that a single VIP is enough to unveil a substantial part of L7LBs for a given PoP.
Given that homogeneous configurations across PoPs are common because they ease maintenance, we argue that not observing all PoPs is not a short-coming. 
Additionally, our findings are confirmed through active measurements, analysis of flow records, and by prior work~\cite{zbsja-io9ae-21}. 
Based on our complementary active measurements, analysis of flow data, and discussions with the operator community, we are confident that insights resulting from backscatter traffic are~valid.

\paragraphNoDot{Will deployment of structured connection IDs increase? Can we apply the same methods to other deployments?}
Structured CIDs may serve as a fingerprint of specific hypergiants.
We find that Google migrated to such connection IDs in 2023, and we observe distinct information encoding from Akamai, Amazon, Apple, Cloudflare, Fastly, Meta, and Microsoft in backscatter.
We argue that the usage will increase over time since they simplify fine-grained provider controlled routing, but standardization might limit the cardinality of unique identification properties and as such our method. 
Advanced QUIC features such as client migration even require additional data encoded in such IDs to prevent overhead from synchronizing connection state.

Our detection of off-net deployments is applicable to other deployments and measurement methods such as flow records without ground-truth knowledge from open source implementations, while detection of layer 7 loadbalancers is limited to Meta, since only Meta uses a cleartext encoding. 

\paragraphNoDot{What are the implications from knowing the number of loadbalancers?}
Encoding the destination loadbalancer into a connection ID enables clients to steer traffic to specific loadbalancer instances. 
This is unwanted behavior because attackers could direct traffic to a single loadbalancer, bypassing single point of failure mitigation.
Although the number of loadbalancer instances does not reveal the underlying capacity and compute power, knowing the distribution in a geographic region or size of a single cluster can be used to estimate the load necessary to overload that PoP.
This information is not only valuable for attackers but also for competitors, allowing them to \one  anticipate business opportunities and local competition, \two the importance of a region, and \three improving their own infrastructure.

%% file: text/conclusion.tex
\section{Conclusion and Outlook}
\label{sec:conclusion}

In this paper, we showed how passive backscatter traffic recorded at the Internet edge can be utilized to closely monitor the rollout of QUIC and in particular reveal deployed configurations of local QUIC stacks and larger distribution infrastructures.
We explicitly quantify layer 7 loadbalancers, which were previously disguised behind virtual IP addresses.

Since passive measurements are completely uncontrolled, \ie based on data triggered by third parties, we used alternative means---captured flow data and active scans---to validate our findings.

\paragraph{Future Trends based on Common Protocols}
Scans reveal that hypergiants operate between 1.6$\times$ and 2.1$\times$ (Cloudflare, Meta), up to 7.5$\times$ (Google) as many TLS/TCP endpoints by number of VIPs than QUIC, but the number of QUIC endpoints is rapidly increasing.
We observe 2.5$\times$ (Meta) to 3$\times$ (Google) the amount of QUIC endpoints in 2025 than in 2021. 
The Cloudflare deployment increased by 17\% during this time.
A continued rise in endpoints suggests a potential for higher victim counts in the future.
Understanding the reasons for not seeing significant backscatter from all QUIC-enabled hypergiants (\eg fewer attacks or filtering) will be part of our future work.
We argue that the proposed approach will gain precision with increasing QUIC~deployment.

%% file: text/acks.tex
\begin{acks}
This work was partly supported by the \grantsponsor{BMFTR}{Federal Ministry of Research, Technology and Space (BMFTR)}{https://www.bmbf.de/} within the project AI.Auto-Immune (\grantnum{BMFTR}{16KIS2333} and \grantnum{BMFTR}{16KIS2332K}).
\end{acks}

%% file: text/appendix.tex
\section{Ethics}
Our proposed method aims for less intrusive measurements.

\paragraph{Backscatter from Network Telescope}
We introduce passive measurements capturing unsolicited traffic via a network telescope. 
The telescope may receive sensitive information, \eg backscatter from victims of source spoofed attacks and compromised hosts.
Unanonymized data is only processed in a controlled environment, and we only share aggregated statistics and our analysis.

\paragraph{NREN Flow Records}
The privacy of NREN users is utmost important to us. 
All work based on NREN data was conducted with great care.
We want to emphasize that the flow data we received did not include full client IP addresses but was already stripped.
Therefore, it is not possible to trace the identity of the data subjects.
Moreover, the dataset consists solely of flow records; no PCAP files were included, and we have never had access to payload data, apart from metadata available in QUIC handshakes.
The NREN legitimately obtained this data on the basis of a legitimate interest (i.e., not consent), among other things, for the purpose of ensuring the further development of services provided to the scientific and research community (which is also why the data has been provided to us).

\paragraph{Active Measurements}
All active measurements conducted in this study complement or verify our results using common techniques and adhere to best practices.
To reduce potential conflicts, we limit the amount of active scanning.

\section{Artifacts}
\label{app:artifcats}

Due to agreements with the data providers we cannot publish telescope data. 
To facilitate further research, we publish the QUIC flows sampled from CESNET and our analysis. %

\paragraphNoDot{Data set:}
QUIC flow records, active measurements of Meta on-net deployments

\paragraphNoDot{Run-time environment:} 
Python, Jupyter Notebooks

\paragraphNoDot{How much disk space is required?} 
\textasciitilde 100 GB

\paragraphNoDot{How much memory is required?}
256~GB

\paragraphNoDot{Publicly available?}
Yes

\paragraphNoDot{Archived?} 
Yes. \\
Analysis: \url{https://doi.org/10.5281/zenodo.17232715} \\
CESNET flow data: \url{https://doi.org/10.5281/zenodo.17249078} \\

\section{Changes in the UCSD Network Telescope}
\label{sec:ucsd-changes}

\begin{figure}[H]
    \centering
    \begin{minipage}[t]{0.48\textwidth}
      \begin{center}
        \includegraphics[width=\textwidth]{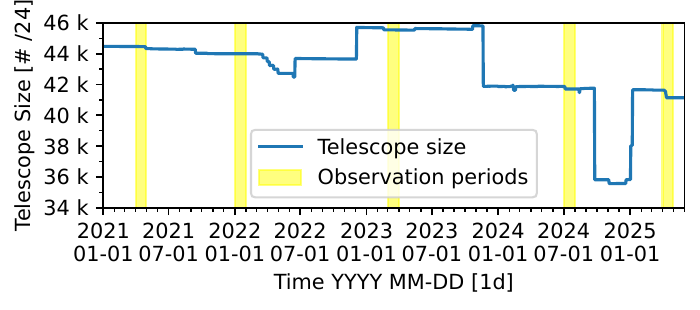}
      \end{center}
      \caption{Size of the UCSD network telescope and observation periods. The size of the UCSD network telescope decreases within our measurement period.}
      \label{fig:ucsd-size}
    \end{minipage}
    \hfill
    \begin{minipage}[t]{0.48\textwidth}
      \begin{center}
        \includegraphics[width=\textwidth]{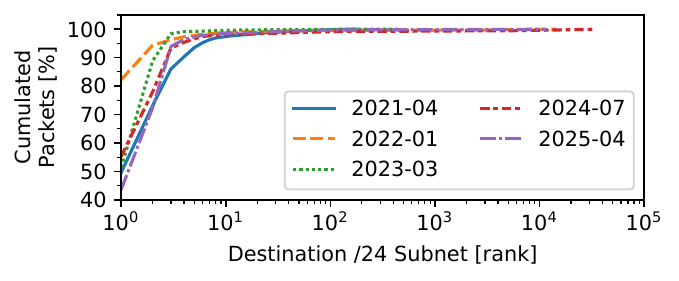}
      \end{center}
      \caption{
        Cumulated reception of QUIC backscatter packets by /24 telescope subnet. 
        The majority of QUIC backscatter is received by only a few subnets.
    }
      \label{fig:cum-packets-per-rank-24}
    \end{minipage}
    \vspace{-1em}
\end{figure}

The address space of the UCSD network telescope, located in a /9 and /10 IPv4 prefix, changes due to legitimate use by the owners of the address space~\cite{mmcgm-lloln-25}. 
\autoref{fig:ucsd-size} shows the size of the UCSD network telescope decreases throughout our measurement period and highlights our telescope observation periods. 
At the end of our observation period, the telescope is~\textasciitilde6\% smaller than in our first observation period. 
In 2023,~\textasciitilde2~\% more addresses are part of the telescope compared to 2021.
The maximum difference equals the size of a \textasciitilde/20 subnet.
Due to missing ground-truth it is impossible to determine how that influences telescope traffic.
\autoref{fig:cum-packets-per-rank-24} shows the dominance of a few target /24 subnets in backscatter. 
Within each dataset, a single /24 subnet contributes $>$43\% of the backscatter traffic.
They contain repeated octets.
Not all subnets with large shares of traffic in one dataset are part of the telescope throughout our measurement period. 
However, the two most active subnets are present throughout our observation period. 
We are unable to dissect how this affects the overall dataset and amount of QUIC backscatter received.

The UCSD network telescope can experience dataloss~\cite{mmcgm-lloln-25}. 
Our dataset it affected by this, but we are unable to determine to which extend more QUIC backscatter could have been received.
However, our method is based on the volume of received QUIC packets, not requiring reception of specific backscatter.

\section{Details on Meta Connection IDs}
\label{sec:facebbok-connectionids}
The Meta QUIC implementation mvfst, encodes information in CIDs. 
The meaning of the encoding is provided in their public source code~\cite{i-m-ef-19}.
\autoref{tab:facebook-scid-encoding} presents details about the encoding schema.

\begin{table}[h]
\caption{Meta SCID structure includes clear text host IDs and reduces randomness to less than half of the SCID length.}
  \label{tab:facebook-scid-encoding}
\begin{tabular}{cccccc}
\toprule
\multicolumn{1}{l}{\textbf{}} & \multicolumn{5}{c}{\textbf{Bits of the SCID}}                                  \\ 
\cmidrule{2-6}
\textbf{SCID Version}         & {Version} & {Host ID} & {Worker ID} &{Process ID} & {Remaining random bits} \\ 
\midrule
1                             & 0-1              & 2-17             & 18-25              & 26                  & 27-63                             \\
2                             & 0-1              & 8-31             & 32-39              & 40                  & 2-7,41-63                             \\
3                             & 0-1              & 8-39             & 40-47              & 48                  & 2-7,49-63                             \\
\bottomrule
\end{tabular}
\end{table}
          
\section{Information encoding in the QUIC header}
\label{sec:information-encoding}

QUIC tries to hide metadata but \textit{Initial} and \textit{Handshake} messages carry data about the deployed QUIC stack.
This includes in cleartext the QUIC version a client offers and a server accepts, the destination and source connection IDs, the retry token and the packet Length information.
Only type-specific nits, packet number length, and the packet number are protected by header protections keys.
The respective keys are derived using a version specific salt and the client DCID set by the client in its first flight for  \textit{Initial} packets and using asymmetric cryptography for all other headers.
As such the header protection keys cannot be calculated in backscatter, because we lack the DCID from the client \textit{Initial} or key material. 

\autoref{tab:quic-metadata} lists all QUIC header fields, whether they are protected, and other limitations for successful extraction.

\begin{table}[h]
\caption{QUIC metadata in long (Initial, Handshake, Version Negotiation, 0-RTT, Retry) and short header packets (1-RTT).}
\label{tab:quic-metadata}
\begin{minipage}{\columnwidth}
  \centering
\begin{tabular}{lcc}
\toprule
                                                                                                                                 & \multicolumn{2}{c}{\textbf{Protected in}}       \\ 
                                                                                                                                 \cmidrule{2-3} 
\textbf{Field}                                                                                                                   & Long Header            & Short Header           \\ 
\cmidrule{1-3}
Header Form                                                                                                                      & \xmark  & \xmark  \\
Fixed Bit                                                                                                                        & \xmark & \xmark \\
Long Packet Type                                                                                                            & \xmark  & -  \\
\begin{tabular}[c]{@{}l@{}}QUIC Type-Specific Bits \\ (Reserved Bits, \\ Spin Bit, Key Phase)\end{tabular} & \cmark  & \cmark  \\
Version                                                                                                                          & \xmark  & -  \\
DCID Length                                                                                                                      & \xmark  & -                      \\
DCID                                                                                                                             & \xmark  & \xmark$^{\dagger}$  \\
SCID Length                                                                                                                      & \xmark  & -                      \\
SCID                                                                                                                             & \xmark  & -                      \\
Token Length                                                                                                                     & \xmark  & -                      \\
Token                                                                                                                            & \xmark  & -                      \\
Length                                                                                                                           & \xmark  & -                      \\
Packet Number Length  & \cmark & - \\
Packet Number                                                                                                                    & \cmark  & \cmark  \\
\bottomrule
                                                                                                                                 &                        &                       
\end{tabular}
\footnotesize{\center$^{\dagger}$ Without knowledge of the DCID length due to its variable length,\\ the DCID cannot be extracted from short header packets.}
\end{minipage}
\end{table}

\section{Measurement statistics}
\label{sec:backscatter-prospects}

Sanitization removes up to 99.9\% of packets, which mainly consist of research scans targeting the complete telescope; those scans have little significance to our analysis, since research infrastructure is well-documented and does not have to be inferred.%

\paragraph{Backscatter}
\autoref{tab:app-ip-addr-host-ids} lists the number of connections in backscatter in the respective year.
We determine QUIC connections by unique SCID, DCID, source and destination IP addresses, and destination port.

\begin{table}[h]
\small
\caption{Number of QUIC connections in backscatter. 
We observe a strong increase in the number of QUIC connections. 
}
\label{tab:app-ip-addr-host-ids}

\begin{tabular*}{\linewidth}{@{\extracolsep{\fill}}lrrrrrrrrr}
\toprule
 & \multicolumn{9}{c}{QUIC connections from Source Networks[\#]} \\ 
\cmidrule{2-10}
Year & Cloudflare & Google & Meta & Akamai & Amazon & Apple & Fastly & Microsoft & Others \\ 
\midrule\addlinespace[2.5pt]
2021 & 39 & 62,403 & 9,552 & - & 1 & - & - & - & 9,290 \\
2022 & 170 & 113,879 & 63,615 & 15 & 2 & 2 & - & 102 & 47,030 \\
2023 & 1,126 & 314,219 & 257,002 & 5,173 & 3,356 & 108 & 276 & 2,422 & 133,083 \\
2024 & 163,117 & 147,936 & 162,088 & 13,787 & 4,485 & 27,958 & 7,478 & 208 & 160,835 \\
2025 & 68,830 & 160,909 & 231,880 & 14,984 & 3,435 & 35,620 & 9,597 & 707 & 73,645 \\
\bottomrule
\end{tabular*}
\end{table}

\paragraph{NREN Flow Data}
\autoref{tab:flows-stats-vips} shows the number of IP addresses of large content providers in flow records from a European NREN in 2024 and 2025.
\autoref{tab:flows-stats-flows} shows the number of flows of large content providers in the same dataset.

\begin{table}[h]
\small
\setlength{\tabcolsep}{2pt}
\caption{Number of IP addresses in flow records from European NREN in 2024 and 2025.}
\label{tab:flows-stats-vips}
\begin{tabular}{@{\extracolsep{\fill}}lrrrrrrrrr}
\toprule
 & \multicolumn{9}{c}{IP addresses from source network [\#]} \\ 
\cmidrule{2-10}
Year & Akamai & Amazon & Apple & Cloudflare & Fastly & Google & Meta & Microsoft & Other ASes \\ 
\midrule
2024& 1,568 & 1,927 & 116 & 5,425 & 132 & 7,217 & 449 & 464 & 1,363 \\
2025& 3,012 & 1,088 & 115 & 8,093 & 184 & 7,478 & 698 & 1,389 & 2,673 \\
\bottomrule
\end{tabular}
\end{table}

\begin{table}[h]
\small
\setlength{\tabcolsep}{2pt}
\caption{Number of flows in flow records from European NREN in 2024 and 2025.}
\label{tab:flows-stats-flows}
\begin{tabular}{@{\extracolsep{\fill}}lrrrrrrrrr}
\toprule
 & \multicolumn{9}{c}{Flows from Source Network [\#]} \\ 
\cmidrule(lr){2-10}
Year & Akamai & Amazon & Apple & Cloudflare & Fastly & Google & Meta & Microsoft & Others ASes \\ 
\midrule
2024 & 80,081 & 37,115 & 1,603 & 176,460 & 79,280 & 7,470,552 & 606,561 & 115,381 & 46,339 \\
2025 & 418,531 & 55,642 & 3,663 & 378,468 & 299,016 & 9,352,327 & 2,535,108 & 284,153 & 110,270 \\
\bottomrule
\end{tabular}
\end{table}

\section{Off-net Classifiers}
\label{sec:classifiers}

Aside from SCID structure, on-net deployments differ in configured retransmission intervals, packet lengths, and enabled packet coalescence.
Subsequently, we detail these classifiers and show their performance in backscatter.
\autoref{tab:facebook-detection} presents true positive rate, false positive rate \etc of identifying Google and Meta off-net servers.
We do not show classifier performance for Cloudflare, since none of the potential off-net candidates allowed for QUIC connections at the time of our measurement.
The bad performance of the Retransmission Interval classifier for Google originates from the less consistent configuration, \ie not all connections have 3 retransmissions.

\paragraph{Retransmission Intervals}
The Retransmission Interval classifiers calculate the time since the reception of the first packet of a connection.
We then use the determined value of the Initial RTO and perform binary exponential backoff on this value. 
If the calculated value is larger than 0.4s, we tolerate packets within $\pm$0.1s of that value. 
Otherwise, the set of timeouts overlaps with the timeouts used by other hypergiants, which increases the false positive rate.
We classify an IP address to be operated by a given hypergiant if at least 3 of the retransmissions from distinct backoff intervals are observed in any flow from that IP address.

\paragraph{Packet Length(s)}
The packet length classifier considers an IP address to be operated by a given hypergiant if any of the Top 5 packet length(s) (see \autoref{fig:CDN-packet-length-observation}) is contained in any packet from that address.

\paragraph{Packet Coalescence}
An IP address is detected as coalescing if a coalesced packet originates from that address.
\begin{table}[h]
\small
\setlength{\tabcolsep}{2pt}
\caption{
Performance of classifying off-net Google and Meta servers based on backscatter traffic with the largest set of off-net candidates (2023). IP addresses were assigned to content providers using subject alternative names in certificates collected with QScanner.}
\label{tab:facebook-detection}
\begin{tabular}{lrrrrrrr}
\toprule
Classifier & TPR & FPR & TNR & FNR & Precision & Recall & F-score \\ 
\midrule\addlinespace[2.5pt]
Meta \\
\quad  Retransmission Intervals & 0.857 & 0.266 & 0.734 & 0.143 & 0.596 & 0.857 & 0.703 \\
\quad  Retransmission Intervals \&  SCIDv1 & 0.857 & 0.023 & 0.977 & 0.143 & 0.944 & 0.857 & 0.899 \\
\quad  Retransmission Intervals \&  SCIDv1 \& No Coalescence & 0.857 & 0.021 & 0.979 & 0.143 & 0.949 & 0.857 & 0.901 \\
\quad  Packet Length & 0.995 & 0.117 & 0.883 & 0.005 & 0.796 & 0.995 & 0.885 \\
\quad  Packet Length \&  SCIDv1 \& No Coalescence & 0.995 & 0.033 & 0.967 & 0.005 & 0.932 & 0.995 & 0.963 \\
\quad No Coalescence & 1.0 & 0.136 & 0.864 & 0.0 & 0.771 & 1.0 & 0.871 \\
\quad  SCIDv1 & 1.0 & 0.053 & 0.947 & 0.0 & 0.895 & 1.0 & 0.945 \\
\quad  SCIDv1 \& No Coalescence & 1.0 & 0.044 & 0.956 & 0.0 & 0.912 & 1.0 & 0.954 \\
\quad  Off-net SCIDv1 & 1.0 & 0.015 & 0.985 & 0.0 & 0.967 & 1.0 & 0.983 \\
\midrule
Google \\
\quad  Retransmission Intervals \&  SCIDv1 & 0.374 & 0.006 & 0.994 & 0.626 & 0.951 & 0.374 & 0.536 \\
\quad  Retransmission Intervals \&  SCIDv1 \& Coalescence & 0.374 & 0.006 & 0.994 & 0.626 & 0.951 & 0.374 & 0.536 \\
\quad  Retransmission Intervals & 0.384 & 0.217 & 0.783 & 0.616 & 0.338 & 0.384 & 0.36 \\
\quad  Packet Length \&  SCIDv1 \& Coalescence & 0.896 & 0.024 & 0.976 & 0.104 & 0.916 & 0.896 & 0.906 \\
\quad  Packet Length & 0.909 & 0.181 & 0.819 & 0.091 & 0.59 & 0.909 & 0.716 \\
\quad  SCIDv1 \& Coalescence & 0.961 & 0.05 & 0.95 & 0.039 & 0.848 & 0.961 & 0.901 \\
\quad Coalescence & 0.972 & 0.345 & 0.655 & 0.028 & 0.448 & 0.972 & 0.613 \\
\quad  SCIDv1 & 0.985 & 0.062 & 0.938 & 0.015 & 0.82 & 0.985 & 0.895 \\
\bottomrule
\end{tabular}
\end{table}

%% file: main.bbl

\begin{thebibliography}{68}


\ifx \showCODEN    \undefined \def \showCODEN     #1{\unskip}     \fi
\ifx \showDOI      \undefined \def \showDOI       #1{#1}\fi
\ifx \showISBNx    \undefined \def \showISBNx     #1{\unskip}     \fi
\ifx \showISBNxiii \undefined \def \showISBNxiii  #1{\unskip}     \fi
\ifx \showISSN     \undefined \def \showISSN      #1{\unskip}     \fi
\ifx \showLCCN     \undefined \def \showLCCN      #1{\unskip}     \fi
\ifx \shownote     \undefined \def \shownote      #1{#1}          \fi
\ifx \showarticletitle \undefined \def \showarticletitle #1{#1}   \fi
\ifx \showURL      \undefined \def \showURL       {\relax}        \fi
\providecommand\bibfield[2]{#2}
\providecommand\bibinfo[2]{#2}
\providecommand\natexlab[1]{#1}
\providecommand\showeprint[2][]{arXiv:#2}

\bibitem[Ager et~al\mbox{.}(2011)]%
        {amsu-wcc-11}
\bibfield{author}{\bibinfo{person}{Bernhard Ager}, \bibinfo{person}{Wolfgang M{\"u}hlbauer}, \bibinfo{person}{Georgios Smaragdakis}, {and} \bibinfo{person}{Steve Uhlig}.} \bibinfo{year}{2011}\natexlab{}.
\newblock \showarticletitle{{Web Content Cartography}}. In \bibinfo{booktitle}{\emph{Proc. of ACM IMC}}. \bibinfo{publisher}{ACM}, \bibinfo{address}{New York, NY, USA}, \bibinfo{pages}{585--600}.
\newblock


\bibitem[Akbari et~al\mbox{.}(2021)]%
        {asvlb-lbctc-21}
\bibfield{author}{\bibinfo{person}{Iman Akbari}, \bibinfo{person}{Mohammad~A. Salahuddin}, \bibinfo{person}{Leni Ven}, \bibinfo{person}{Noura Limam}, \bibinfo{person}{Raouf Boutaba}, \bibinfo{person}{Bertrand Mathieu}, \bibinfo{person}{Stephanie Moteau}, {and} \bibinfo{person}{Stephane Tuffin}.} \bibinfo{year}{2021}\natexlab{}.
\newblock \showarticletitle{A Look Behind the Curtain: Traffic Classification in an Increasingly Encrypted Web}.
\newblock \bibinfo{journal}{\emph{Proc. ACM Meas. Anal. Comput. Syst.}} \bibinfo{volume}{5}, \bibinfo{number}{1}, Article \bibinfo{articleno}{04} (\bibinfo{date}{feb} \bibinfo{year}{2021}), \bibinfo{numpages}{26}~pages.
\newblock
\urldef\tempurl%
\url{https://doi.org/10.1145/3447382}
\showDOI{\tempurl}


\bibitem[Augustin et~al\mbox{.}(2006)]%
        {acovf-atawp-06}
\bibfield{author}{\bibinfo{person}{Brice Augustin}, \bibinfo{person}{Xavier Cuvellier}, \bibinfo{person}{Benjamin Orgogozo}, \bibinfo{person}{Fabien Viger}, \bibinfo{person}{Timur Friedman}, \bibinfo{person}{Matthieu Latapy}, \bibinfo{person}{Cl\'{e}mence Magnien}, {and} \bibinfo{person}{Renata Teixeira}.} \bibinfo{year}{2006}\natexlab{}.
\newblock \showarticletitle{{Avoiding traceroute anomalies with Paris traceroute}}. In \bibinfo{booktitle}{\emph{Proc. of ACM IMC}} (Rio de Janeriro, Brazil). \bibinfo{publisher}{ACM}, \bibinfo{address}{New York, NY, USA}, \bibinfo{pages}{153--158}.
\newblock


\bibitem[Barbette et~al\mbox{.}(2020)]%
        {btykm-hslbd-20}
\bibfield{author}{\bibinfo{person}{Tom Barbette}, \bibinfo{person}{Chen Tang}, \bibinfo{person}{Haoran Yao}, \bibinfo{person}{Dejan Kostic}, \bibinfo{person}{Gerald Q.~Maguire Jr.}, \bibinfo{person}{Panagiotis Papadimitratos}, {and} \bibinfo{person}{Marco Chiesa}.} \bibinfo{year}{2020}\natexlab{}.
\newblock \showarticletitle{A High-Speed Load-Balancer Design with Guaranteed Per-Connection-Consistency}. In \bibinfo{booktitle}{\emph{17th {USENIX} Symposium on Networked Systems Design and Implementation, {NSDI} 2020, Santa Clara, CA, USA, February 25-27, 2020}}, \bibfield{editor}{\bibinfo{person}{Ranjita Bhagwan} {and} \bibinfo{person}{George Porter}} (Eds.). \bibinfo{publisher}{{USENIX} Association}, \bibinfo{pages}{667--683}.
\newblock
\urldef\tempurl%
\url{https://www.usenix.org/conference/nsdi20/presentation/barbette}
\showURL{%
\tempurl}


\bibitem[Bishop(2022)]%
        {RFC-9114}
\bibfield{author}{\bibinfo{person}{M. Bishop}.} \bibinfo{year}{2022}\natexlab{}.
\newblock \bibinfo{booktitle}{\emph{{HTTP/3}}}.
\newblock \bibinfo{type}{RFC} 9114. \bibinfo{institution}{IETF}.
\newblock
\urldef\tempurl%
\url{https://doi.org/10.17487/RFC9114}
\showDOI{\tempurl}


\bibitem[B{\"o}ttger et~al\mbox{.}(2018)]%
        {bctcu-ocega-18}
\bibfield{author}{\bibinfo{person}{Timm B{\"o}ttger}, \bibinfo{person}{Felix Cuadrado}, \bibinfo{person}{Gareth Tyson}, \bibinfo{person}{Ignacio Castro}, {and} \bibinfo{person}{Steve Uhlig}.} \bibinfo{year}{2018}\natexlab{}.
\newblock \showarticletitle{{Open Connect Everywhere: A Glimpse at the Internet Ecosystem through the Lens of the Netflix CDN}}.
\newblock \bibinfo{journal}{\emph{SIGCOMM Comput. Commun. Rev.}} \bibinfo{volume}{48}, \bibinfo{number}{1} (\bibinfo{date}{apr} \bibinfo{year}{2018}), \bibinfo{pages}{28--34}.
\newblock
\urldef\tempurl%
\url{https://doi.org/10.1145/3211852.3211857}
\showDOI{\tempurl}


\bibitem[CAIDA(2012)]%
        {c-unt-12}
\bibfield{author}{\bibinfo{person}{CAIDA}.} \bibinfo{year}{2012}\natexlab{}.
\newblock \bibinfo{title}{{The UCSD Network Telescope}}.
\newblock \bibinfo{howpublished}{Website}.
\newblock
\urldef\tempurl%
\url{https://www.caida.org/projects/network_telescope/}
\showURL{%
\tempurl}
\newblock
\shownote{Last Access: May 2025}.


\bibitem[CAIDA(2022)]%
        {c-sp-22}
\bibfield{author}{\bibinfo{person}{CAIDA}.} \bibinfo{year}{2022}\natexlab{}.
\newblock \bibinfo{title}{Spoofer Project}.
\newblock
\newblock
\urldef\tempurl%
\url{https://www.caida.org/projects/spoofer/}
\showURL{%
\tempurl}
\newblock
\shownote{Last Access: June 2025}.


\bibitem[CESNET(2019)]%
        {c-ifeds-19}
\bibfield{author}{\bibinfo{person}{CESNET}.} \bibinfo{year}{2019}\natexlab{}.
\newblock \bibinfo{title}{ipfixprobe -- IPFIX flow exporter with DPDK support}.
\newblock \bibinfo{howpublished}{GitHub Repository}.
\newblock
\urldef\tempurl%
\url{https://github.com/CESNET/ipfixprobe}
\showURL{%
\tempurl}
\newblock
\shownote{Last Access: September 2025}.


\bibitem[Cloudflare(2018)]%
        {c-q-18}
\bibfield{author}{\bibinfo{person}{Cloudflare}.} \bibinfo{year}{2018}\natexlab{}.
\newblock \bibinfo{title}{QUICHE}.
\newblock \bibinfo{howpublished}{Github Repository}.
\newblock
\urldef\tempurl%
\url{https://github.com/cloudflare/quiche}
\showURL{%
\tempurl}
\newblock
\shownote{Last Access: Jun 2025}.


\bibitem[Cloudflare(2025a)]%
        {c-crau-25}
\bibfield{author}{\bibinfo{person}{Cloudflare}.} \bibinfo{year}{2025}\natexlab{a}.
\newblock \bibinfo{title}{{Cloudflare Radar: Adoption \& Usage}}.
\newblock
\newblock
\urldef\tempurl%
\url{https://radar.cloudflare.com/adoption-and-usage?dateRange=52w}
\showURL{%
\tempurl}
\newblock
\shownote{Last Access: June 2025}.


\bibitem[Cloudflare(2025b)]%
        {c-ugrm-25}
\bibfield{author}{\bibinfo{person}{Cloudflare}.} \bibinfo{year}{2025}\natexlab{b}.
\newblock \bibinfo{title}{UDP Graceful Restart Marshal}.
\newblock \bibinfo{howpublished}{GitHub Repository}.
\newblock
\urldef\tempurl%
\url{https://github.com/cloudflare/udpgrm/tree/main}
\showURL{%
\tempurl}
\newblock
\shownote{Last Access: June 2025}.


\bibitem[Collins(2021)]%
        {c-as-21}
\bibfield{author}{\bibinfo{person}{Michael Collins}.} \bibinfo{year}{2021}\natexlab{}.
\newblock \bibinfo{title}{Acknowledged Scanners}.
\newblock \bibinfo{howpublished}{GitLab Repository}.
\newblock
\urldef\tempurl%
\url{https://gitlab.com/mcollins_at_isi/acknowledged_scanners}
\showURL{%
\tempurl}
\newblock
\shownote{Last Access: May 2025}.


\bibitem[Cui et~al\mbox{.}(2017)]%
        {cllwk-itwqd-17}
\bibfield{author}{\bibinfo{person}{Yong Cui}, \bibinfo{person}{Tianxiang Li}, \bibinfo{person}{Cong Liu}, \bibinfo{person}{Xingwei Wang}, {and} \bibinfo{person}{Mirja K{ \"{u}}hlewind}.} \bibinfo{year}{2017}\natexlab{}.
\newblock \showarticletitle{{Innovating Transport with {QUIC:} Design Approaches and Research Challenges}}.
\newblock \bibinfo{journal}{\emph{{IEEE} Internet Comput.}} \bibinfo{volume}{21}, \bibinfo{number}{2} (\bibinfo{year}{2017}), \bibinfo{pages}{72--76}.
\newblock
\urldef\tempurl%
\url{https://doi.org/10.1109/MIC.2017.44}
\showDOI{\tempurl}


\bibitem[DeLong(2022)]%
        {d-sifv-22}
\bibfield{author}{\bibinfo{person}{Owen DeLong}.} \bibinfo{year}{2022}\natexlab{}.
\newblock \bibinfo{title}{{Re: Scanning the Internet for Vulnerabilities}}.
\newblock
\newblock
\urldef\tempurl%
\url{https://seclists.org/nanog/2022/Jun/269}
\showURL{%
\tempurl}
\newblock
\shownote{Last Access: June 2025}.


\bibitem[Doan et~al\mbox{.}(2022)]%
        {drhb-evoco-22}
\bibfield{author}{\bibinfo{person}{Trinh~Viet Doan}, \bibinfo{person}{Roland van Rijswijk-Deij}, \bibinfo{person}{Oliver Hohlfeld}, {and} \bibinfo{person}{Vaibhav Bajpai}.} \bibinfo{year}{2022}\natexlab{}.
\newblock \showarticletitle{{An Empirical View on Consolidation of the Web}}.
\newblock \bibinfo{journal}{\emph{ACM Trans. Internet Technol.}} \bibinfo{volume}{22}, \bibinfo{number}{3} (\bibinfo{date}{feb} \bibinfo{year}{2022}), \bibinfo{numpages}{30}~pages.
\newblock
\urldef\tempurl%
\url{https://doi.org/10.1145/3503158}
\showDOI{\tempurl}


\bibitem[Duke et~al\mbox{.}(2024)]%
        {draft-ietf-quic-load-balancers-18}
\bibfield{author}{\bibinfo{person}{Martin Duke}, \bibinfo{person}{Nick Banks}, {and} \bibinfo{person}{Christian Huitema}.} \bibinfo{year}{2024}\natexlab{}.
\newblock \bibinfo{booktitle}{\emph{{QUIC-LB: Generating Routable QUIC Connection IDs}}}.
\newblock \bibinfo{type}{Internet-Draft -- work in progress}~18. \bibinfo{institution}{IETF}.
\newblock
\urldef\tempurl%
\url{https://datatracker.ietf.org/doc/html/draft-ietf-quic-load-balancers-18}
\showURL{%
\tempurl}


\bibitem[Duke et~al\mbox{.}(2025)]%
        {draft-ietf-quic-load-balancers}
\bibfield{author}{\bibinfo{person}{Martin Duke}, \bibinfo{person}{Nick Banks}, {and} \bibinfo{person}{Christian Huitema}.} \bibinfo{year}{2025}\natexlab{}.
\newblock \bibinfo{booktitle}{\emph{{QUIC-LB: Generating Routable QUIC Connection IDs}}}.
\newblock \bibinfo{type}{Internet-Draft -- work in progress}~21. \bibinfo{institution}{IETF}.
\newblock
\urldef\tempurl%
\url{https://datatracker.ietf.org/doc/html/draft-ietf-quic-load-balancers-21}
\showURL{%
\tempurl}


\bibitem[Durumeric et~al\mbox{.}(2024)]%
        {daswh-tyz-24}
\bibfield{author}{\bibinfo{person}{Zakir Durumeric}, \bibinfo{person}{David Adrian}, \bibinfo{person}{Phillip Stephens}, \bibinfo{person}{Eric Wustrow}, {and} \bibinfo{person}{J.~Alex Halderman}.} \bibinfo{year}{2024}\natexlab{}.
\newblock \showarticletitle{{Ten Years of ZMap}}. In \bibinfo{booktitle}{\emph{Proc. of ACM IMC}}. \bibinfo{publisher}{ACM}, \bibinfo{address}{New York, NY, USA}, \bibinfo{pages}{139--148}.
\newblock
\urldef\tempurl%
\url{https://doi.org/10.1145/3646547.3689012}
\showDOI{\tempurl}


\bibitem[Eddy(2007)]%
        {RFC-4987}
\bibfield{author}{\bibinfo{person}{W. Eddy}.} \bibinfo{year}{2007}\natexlab{}.
\newblock \bibinfo{booktitle}{\emph{{TCP SYN Flooding Attacks and Common Mitigations}}}.
\newblock \bibinfo{type}{RFC} 4987. \bibinfo{institution}{IETF}.
\newblock
\urldef\tempurl%
\url{https://doi.org/10.17487/RFC4987}
\showDOI{\tempurl}


\bibitem[Eisenbud et~al\mbox{.}(2016)]%
        {eycsk-mfars-16}
\bibfield{author}{\bibinfo{person}{Daniel~E. Eisenbud}, \bibinfo{person}{Cheng Yi}, \bibinfo{person}{Carlo Contavalli}, \bibinfo{person}{Cody Smith}, \bibinfo{person}{Roman Kononov}, \bibinfo{person}{Eric Mann-Hielscher}, \bibinfo{person}{Ardas Cilingiroglu}, \bibinfo{person}{Bin Cheyney}, \bibinfo{person}{Wentao Shang}, {and} \bibinfo{person}{Jinnah~Dylan Hosein}.} \bibinfo{year}{2016}\natexlab{}.
\newblock \showarticletitle{Maglev: A Fast and Reliable Software Network Load Balancer}. In \bibinfo{booktitle}{\emph{Proc. of USENIX NSDI}}. \bibinfo{publisher}{USENIX Association}, \bibinfo{address}{Santa Clara, CA}, \bibinfo{pages}{523--535}.
\newblock
\urldef\tempurl%
\url{https://www.usenix.org/conference/nsdi16/technical-sessions/presentation/eisenbud}
\showURL{%
\tempurl}


\bibitem[Fayed et~al\mbox{.}(2021)]%
        {fbgkm-ttudi-21}
\bibfield{author}{\bibinfo{person}{Marwan Fayed}, \bibinfo{person}{Lorenz Bauer}, \bibinfo{person}{Vasileios Giotsas}, \bibinfo{person}{Sami Kerola}, \bibinfo{person}{Marek Majkowski}, \bibinfo{person}{Pavel Odintsov}, \bibinfo{person}{Jakub Sitnicki}, \bibinfo{person}{Taejoong Chung}, \bibinfo{person}{Dave Levin}, \bibinfo{person}{Alan Mislove}, \bibinfo{person}{Christopher~A. Wood}, {and} \bibinfo{person}{Nick Sullivan}.} \bibinfo{year}{2021}\natexlab{}.
\newblock \showarticletitle{{The Ties That Un-Bind: Decoupling IP from Web Services and Sockets for Robust Addressing Agility at CDN-Scale}}. In \bibinfo{booktitle}{\emph{Proc. of ACM SIGCOMM}}. \bibinfo{publisher}{ACM}, \bibinfo{address}{New York, NY, USA}, \bibinfo{pages}{433--446}.
\newblock
\urldef\tempurl%
\url{https://doi.org/10.1145/3452296.3472922}
\showDOI{\tempurl}


\bibitem[Gigis et~al\mbox{.}(2021)]%
        {gcmnk-syilo-21}
\bibfield{author}{\bibinfo{person}{Petros Gigis}, \bibinfo{person}{Matt Calder}, \bibinfo{person}{Lefteris Manassakis}, \bibinfo{person}{George Nomikos}, \bibinfo{person}{Vasileios Kotronis}, \bibinfo{person}{Xenofontas Dimitropoulos}, \bibinfo{person}{Ethan Katz-Bassett}, {and} \bibinfo{person}{Georgios Smaragdakis}.} \bibinfo{year}{2021}\natexlab{}.
\newblock \showarticletitle{{Seven Years in the Life of Hypergiants' Off-Nets}}. In \bibinfo{booktitle}{\emph{Proc. of ACM SIGCOMM}}. \bibinfo{publisher}{ACM}, \bibinfo{address}{New York, NY, USA}, \bibinfo{pages}{516--533}.
\newblock
\urldef\tempurl%
\url{https://doi.org/10.1145/3452296.3472928}
\showURL{%
\tempurl}


\bibitem[Google(2021)]%
        {g-q-21}
\bibfield{author}{\bibinfo{person}{Google}.} \bibinfo{year}{2021}\natexlab{}.
\newblock \bibinfo{title}{QUICHE}.
\newblock \bibinfo{howpublished}{GitHub Repository}.
\newblock
\urldef\tempurl%
\url{https://github.com/google/quiche}
\showURL{%
\tempurl}
\newblock
\shownote{Last Access: Jul 2025}.


\bibitem[Guilmette et~al\mbox{.}(2022)]%
        {g-sifv-22}
\bibfield{author}{\bibinfo{person}{Ronald~F. Guilmette} {et~al\mbox{.}}} \bibinfo{year}{2022}\natexlab{}.
\newblock \bibinfo{title}{{Scanning the Internet for Vulnerabilities}}.
\newblock
\newblock
\urldef\tempurl%
\url{https://seclists.org/nanog/2022/Jun/250}
\showURL{%
\tempurl}
\newblock
\shownote{Last Access: June 2025}.


\bibitem[Hiesgen et~al\mbox{.}(2024)]%
        {hnbkh-adeci-24}
\bibfield{author}{\bibinfo{person}{Raphael Hiesgen}, \bibinfo{person}{Marcin Nawrocki}, \bibinfo{person}{Marinho Barcellos}, \bibinfo{person}{Daniel Kopp}, \bibinfo{person}{Oliver Hohlfeld}, \bibinfo{person}{Echo Chan}, \bibinfo{person}{Roland Dobbins}, \bibinfo{person}{Christian Doerr}, \bibinfo{person}{Christian Rossow}, \bibinfo{person}{Daniel~R. Thomas}, \bibinfo{person}{Mattijs Jonker}, \bibinfo{person}{Ricky Mok}, \bibinfo{person}{Xiapu Luo}, \bibinfo{person}{John Kristoff}, \bibinfo{person}{Thomas~C. Schmidt}, \bibinfo{person}{Matthias W{\"a}hlisch}, {and} \bibinfo{person}{KC Claffy}.} \bibinfo{year}{2024}\natexlab{}.
\newblock \showarticletitle{{The Age of DDoScovery: An Empirical Comparison of Industry and Academic DDoS Assessments}}. In \bibinfo{booktitle}{\emph{Proc. of ACM Internet Measurement Conference (IMC)}}. \bibinfo{publisher}{ACM}, \bibinfo{address}{New York}, \bibinfo{pages}{259--279}.
\newblock
\urldef\tempurl%
\url{https://doi.org/10.1145/3646547.3688451}
\showURL{%
\tempurl}


\bibitem[Hiesgen et~al\mbox{.}(2022)]%
        {hnkds-sunwo-22}
\bibfield{author}{\bibinfo{person}{Raphael Hiesgen}, \bibinfo{person}{Marcin Nawrocki}, \bibinfo{person}{Alistair King}, \bibinfo{person}{Alberto Dainotti}, \bibinfo{person}{Thomas~C. Schmidt}, {and} \bibinfo{person}{Matthias W{\"a}hlisch}.} \bibinfo{year}{2022}\natexlab{}.
\newblock \showarticletitle{{Spoki: Unveiling a New Wave of Scanners through a Reactive Network Telescope}}. In \bibinfo{booktitle}{\emph{Proc. of USENIX Security}}. \bibinfo{publisher}{USENIX Association}, \bibinfo{address}{Berkeley, CA, USA}, \bibinfo{pages}{431--448}.
\newblock
\urldef\tempurl%
\url{https://www.usenix.org/conference/usenixsecurity22/presentation/hiesgen}
\showURL{%
\tempurl}


\bibitem[Hilal et~al\mbox{.}(2024)]%
        {hsvg-flihi-24}
\bibfield{author}{\bibinfo{person}{Fahad Hilal}, \bibinfo{person}{Patrick Sattler}, \bibinfo{person}{Kevin Vermeulen}, {and} \bibinfo{person}{Oliver Gasser}.} \bibinfo{year}{2024}\natexlab{}.
\newblock \showarticletitle{{A First Look At IPv6 Hypergiant Infrastructure}}.
\newblock \bibinfo{journal}{\emph{Proc. ACM Netw.}} \bibinfo{volume}{2}, \bibinfo{number}{CoNEXT2} (\bibinfo{date}{June} \bibinfo{year}{2024}), \bibinfo{pages}{11:1--11:25}.
\newblock
\urldef\tempurl%
\url{https://doi.org/10.1145/3656300}
\showURL{%
\tempurl}


\bibitem[Hus{\'{a}}k et~al\mbox{.}(2016)]%
        {hcjc-htaac-16}
\bibfield{author}{\bibinfo{person}{Martin Hus{\'{a}}k}, \bibinfo{person}{Milan Cerm{\'{a}}k}, \bibinfo{person}{Tom{\'{a}}s Jirs{\'{\i }}k}, {and} \bibinfo{person}{Pavel Celeda}.} \bibinfo{year}{2016}\natexlab{}.
\newblock \showarticletitle{{HTTPS} traffic analysis and client identification using passive { SSL/TLS} fingerprinting}.
\newblock \bibinfo{journal}{\emph{{EURASIP} J. Inf. Secur.}}  \bibinfo{volume}{2016} (\bibinfo{year}{2016}), \bibinfo{pages}{6}.
\newblock
\urldef\tempurl%
\url{https://doi.org/10.1186/s13635-016-0030-7}
\showDOI{\tempurl}


\bibitem[Incubator(2019a)]%
        {i-m-ef-19-2}
\bibfield{author}{\bibinfo{person}{Facebook Incubator}.} \bibinfo{year}{2019}\natexlab{a}.
\newblock \bibinfo{title}{mvfst}.
\newblock \bibinfo{howpublished}{GitHub Repository}.
\newblock
\urldef\tempurl%
\url{https://github.com/facebookincubator/mvfst}
\showURL{%
\tempurl}
\newblock
\shownote{Last Access: June 2025}.


\bibitem[Incubator(2019b)]%
        {i-m-ef-19}
\bibfield{author}{\bibinfo{person}{Facebook Incubator}.} \bibinfo{year}{2019}\natexlab{b}.
\newblock \bibinfo{title}{mvfst -- encodeConnectionId function}.
\newblock \bibinfo{howpublished}{GitHub Repository}.
\newblock
\urldef\tempurl%
\url{https://github.com/facebookincubator/mvfst/blob/9603981b74a7d28004331e6fd6dbf2882ad2c291/quic/codec/DefaultConnectionIdAlgo.cpp#L323}
\showURL{%
\tempurl}
\newblock
\shownote{Last Access: June 2025}.


\bibitem[Iyengar and Thomson(2021)]%
        {RFC-9000}
\bibfield{author}{\bibinfo{person}{J. Iyengar} {and} \bibinfo{person}{M. Thomson}.} \bibinfo{year}{2021}\natexlab{}.
\newblock \bibinfo{booktitle}{\emph{{QUIC: A UDP-Based Multiplexed and Secure Transport}}}.
\newblock \bibinfo{type}{RFC} 9000. \bibinfo{institution}{IETF}.
\newblock
\urldef\tempurl%
\url{https://doi.org/10.17487/RFC9000}
\showDOI{\tempurl}


\bibitem[Jaeger et~al\mbox{.}(2023)]%
        {jzkpc-qohep-23}
\bibfield{author}{\bibinfo{person}{Benedikt Jaeger}, \bibinfo{person}{Johannes Zirngibl}, \bibinfo{person}{Marcel Kempf}, \bibinfo{person}{Kevin Ploch}, {and} \bibinfo{person}{Georg Carle}.} \bibinfo{year}{2023}\natexlab{}.
\newblock \showarticletitle{QUIC on the Highway:Evaluating Performance on High-rate Links}. In \bibinfo{booktitle}{\emph{In Proc. of IFIP Networking Conference (IFIP Networking)}}. \bibinfo{publisher}{IFIP}, \bibinfo{address}{Barcelona, Spain}.
\newblock


\bibitem[John and Tafvelin(2007)]%
        {jt-aibth-07}
\bibfield{author}{\bibinfo{person}{Wolfgang John} {and} \bibinfo{person}{Sven Tafvelin}.} \bibinfo{year}{2007}\natexlab{}.
\newblock \showarticletitle{Analysis of internet backbone traffic and header anomalies observed}. In \bibinfo{booktitle}{\emph{Proceedings of the 7th ACM SIGCOMM Conference on Internet Measurement}} (San Diego, California, USA) \emph{(\bibinfo{series}{IMC '07})}. \bibinfo{publisher}{Association for Computing Machinery}, \bibinfo{address}{New York, NY, USA}, \bibinfo{pages}{111–116}.
\newblock
\showISBNx{9781595939081}
\urldef\tempurl%
\url{https://doi.org/10.1145/1298306.1298321}
\showDOI{\tempurl}


\bibitem[Jones(2018)]%
        {j-ghswq-18}
\bibfield{author}{\bibinfo{person}{Nick Jones}.} \bibinfo{year}{2018}\natexlab{}.
\newblock \bibinfo{title}{Get a head start with QUIC}.
\newblock \bibinfo{howpublished}{Blog}.
\newblock
\urldef\tempurl%
\url{https://blog.cloudflare.com/head-start-with-quic/}
\showURL{%
\tempurl}
\newblock
\shownote{Last Access: May 2025}.


\bibitem[Joras(2024)]%
        {j-wiqg-24}
\bibfield{author}{\bibinfo{person}{Matt Joras}.} \bibinfo{year}{2024}\natexlab{}.
\newblock \bibinfo{title}{{Re: Why isn't QUIC growing?}}
\newblock
\newblock
\urldef\tempurl%
\url{https://mailarchive.ietf.org/arch/msg/quic/SqMCCWSyVeI46Exu4I-MCAAgg_w/}
\showURL{%
\tempurl}
\newblock
\shownote{Last Access: June 2025}.


\bibitem[Kistenmacher et~al\mbox{.}(2025)]%
        {ktf-qalba-25}
\bibfield{author}{\bibinfo{person}{Liliana Kistenmacher}, \bibinfo{person}{Anum Talpur}, {and} \bibinfo{person}{Mathias Fischer}.} \bibinfo{year}{2025}\natexlab{}.
\newblock \showarticletitle{QUIC-Aware Load Balancing: Attacks and Mitigations}. In \bibinfo{booktitle}{\emph{55th Annual {IEEE/IFIP} International Conference on Dependable Systems and Networks, {DSN} 2025, Naples, Italy, June 23-26, 2025}}. \bibinfo{publisher}{{IEEE}}, \bibinfo{pages}{524--536}.
\newblock
\urldef\tempurl%
\url{https://doi.org/10.1109/DSN64029.2025.00056}
\showDOI{\tempurl}


\bibitem[Koch et~al\mbox{.}(2021)]%
        {kkhca-aicto-21}
\bibfield{author}{\bibinfo{person}{Thomas Koch}, \bibinfo{person}{Ethan Katz-Bassett}, \bibinfo{person}{John Heidemann}, \bibinfo{person}{Matt Calder}, \bibinfo{person}{Calvin Ardi}, {and} \bibinfo{person}{Ke Li}.} \bibinfo{year}{2021}\natexlab{}.
\newblock \showarticletitle{{Anycast In Context: A Tale of Two Systems}}. In \bibinfo{booktitle}{\emph{Proc. of ACM SIGCOMM}}. \bibinfo{publisher}{ACM}, \bibinfo{address}{New York, NY, USA}, \bibinfo{pages}{398--417}.
\newblock
\urldef\tempurl%
\url{https://doi.org/10.1145/3452296.3472891}
\showURL{%
\tempurl}


\bibitem[Labovitz et~al\mbox{.}(2010)]%
        {limoj-iit-10}
\bibfield{author}{\bibinfo{person}{Craig Labovitz}, \bibinfo{person}{Scott Iekel-Johnson}, \bibinfo{person}{Danny McPherson}, \bibinfo{person}{Jon Oberheide}, {and} \bibinfo{person}{Farnam Jahanian}.} \bibinfo{year}{2010}\natexlab{}.
\newblock \showarticletitle{Internet Inter-Domain Traffic}.
\newblock \bibinfo{journal}{\emph{ACM Sigcomm Computer Communication Review}} \bibinfo{volume}{40}, \bibinfo{number}{4} (\bibinfo{date}{Aug} \bibinfo{year}{2010}), \bibinfo{pages}{75--86}.
\newblock
\urldef\tempurl%
\url{https://doi.org/10.1145/1851275.1851194}
\showDOI{\tempurl}


\bibitem[Luxemburk et~al\mbox{.}(2023)]%
        {lhc-etcqc-23}
\bibfield{author}{\bibinfo{person}{Jan Luxemburk}, \bibinfo{person}{Karel Hynek}, {and} \bibinfo{person}{Tomáš Čejka}.} \bibinfo{year}{2023}\natexlab{}.
\newblock \showarticletitle{Encrypted traffic classification: the QUIC case}. In \bibinfo{booktitle}{\emph{Proc. of Network Traffic Measurement and Analysis Conference (TMA)}} (Naples, Italy). \bibinfo{publisher}{IFIP}, \bibinfo{address}{Naples, Italy}.
\newblock


\bibitem[Lyon(2009)]%
        {g-nnson-09}
\bibfield{author}{\bibinfo{person}{Gordon~Fyodor Lyon}.} \bibinfo{year}{2009}\natexlab{}.
\newblock \bibinfo{booktitle}{\emph{{Nmap Network Scanning: The Official Nmap Project Guide to Network Discovery and Security Scanning}}}.
\newblock \bibinfo{publisher}{Nmap Project}, Chapter Legal Issues.
\newblock


\bibitem[Madariaga et~al\mbox{.}(2020)]%
        {mtmbb-aaoqf-20}
\bibfield{author}{\bibinfo{person}{Diego Madariaga}, \bibinfo{person}{Lucas Torrealba}, \bibinfo{person}{Javier Madariaga}, \bibinfo{person}{Javiera Berm~{\'u}dez}, {and} \bibinfo{person}{Javier Bustos-Jim{\'e}nez}.} \bibinfo{year}{2020}\natexlab{}.
\newblock \showarticletitle{Analyzing the Adoption of QUIC From a Mobile Development Perspective}. In \bibinfo{booktitle}{\emph{Proc. of the Workshop on the Evolution, Performance, and Interoperability of QUIC}} (Virtual Event, USA) \emph{(\bibinfo{series}{EPIQ '20})}. \bibinfo{publisher}{{ACM}}, \bibinfo{address}{New York, NY, USA}, \bibinfo{pages}{35--41}.
\newblock
\showISBNx{9781450380478}
\urldef\tempurl%
\url{https://doi.org/10.1145/3405796.3405830}
\showDOI{\tempurl}


\bibitem[M{\"a}nnel et~al\mbox{.}(2025)]%
        {mmcgm-lloln-25}
\bibfield{author}{\bibinfo{person}{Alexander M{\"a}nnel}, \bibinfo{person}{Jonas M{\"u}cke}, \bibinfo{person}{kc Claffy}, \bibinfo{person}{Max Gao}, \bibinfo{person}{Ricky K.~P. Mok}, \bibinfo{person}{Marcin Nawrocki}, \bibinfo{person}{Thomas~C. Schmidt}, {and} \bibinfo{person}{Matthias W{\"a}hlisch}.} \bibinfo{year}{2025}\natexlab{}.
\newblock \showarticletitle{{Lessons Learned from Operating a Large Network Telescope}}. In \bibinfo{booktitle}{\emph{Proc. of ACM Special Interest Group on Data Communication (SIGCOMM)}}. \bibinfo{publisher}{ACM}, \bibinfo{address}{New York}, \bibinfo{pages}{826--841}.
\newblock
\urldef\tempurl%
\url{https://doi.org/10.1145/3718958.3754347}
\showDOI{\tempurl}


\bibitem[Marx et~al\mbox{.}(2020)]%
        {mhlq-ssdds-20}
\bibfield{author}{\bibinfo{person}{Robin Marx}, \bibinfo{person}{Joris Herbots}, \bibinfo{person}{Wim Lamotte}, {and} \bibinfo{person}{Peter Quax}.} \bibinfo{year}{2020}\natexlab{}.
\newblock \showarticletitle{Same Standards, Different Decisions: {A} Study of {QUIC} and {HTTP/3} Implementation Diversity}. In \bibinfo{booktitle}{\emph{Proceedings of the 2020 Workshop on the Evolution, Performance, and Interoperability of QUIC, EPIQ@SIGCOMM 2020, Virtual Event, USA, August 10-14, 2020}}, \bibfield{editor}{\bibinfo{person}{J{\"o}rg Ott} {and} \bibinfo{person}{Lars Eggert}} (Eds.). \bibinfo{publisher}{{ACM}}, \bibinfo{address}{New York, NY, USA}, \bibinfo{pages}{14--20}.
\newblock
\urldef\tempurl%
\url{https://doi.org/10.1145/3405796.3405828}
\showDOI{\tempurl}


\bibitem[Mellia et~al\mbox{.}(2008)]%
        {mmmr-pata-08}
\bibfield{author}{\bibinfo{person}{Marco Mellia}, \bibinfo{person}{Michela Meo}, \bibinfo{person}{Luca Muscariello}, {and} \bibinfo{person}{Dario Rossi}.} \bibinfo{year}{2008}\natexlab{}.
\newblock \showarticletitle{Passive analysis of {TCP} anomalies}.
\newblock \bibinfo{journal}{\emph{Comput. Networks}} \bibinfo{volume}{52}, \bibinfo{number}{14} (\bibinfo{year}{2008}), \bibinfo{pages}{2663--2676}.
\newblock
\urldef\tempurl%
\url{https://doi.org/10.1016/J.COMNET.2008.05.010}
\showDOI{\tempurl}


\bibitem[Minaburo and Toutain(2023)]%
        {RFC-9363}
\bibfield{author}{\bibinfo{person}{A. Minaburo} {and} \bibinfo{person}{L. Toutain}.} \bibinfo{year}{2023}\natexlab{}.
\newblock \bibinfo{booktitle}{\emph{{A YANG Data Model for Static Context Header Compression (SCHC)}}}.
\newblock \bibinfo{type}{RFC} 9363. \bibinfo{institution}{IETF}.
\newblock
\urldef\tempurl%
\url{https://doi.org/10.17487/RFC9363}
\showDOI{\tempurl}


\bibitem[M{\"u}cke et~al\mbox{.}(2024)]%
        {mnhsw-rqmpi-24}
\bibfield{author}{\bibinfo{person}{Jonas M{\"u}cke}, \bibinfo{person}{Marcin Nawrocki}, \bibinfo{person}{Raphael Hiesgen}, \bibinfo{person}{Thomas~C. Schmidt}, {and} \bibinfo{person}{Matthias W{\"a}hlisch}.} \bibinfo{year}{2024}\natexlab{}.
\newblock \showarticletitle{{ReACKed QUICer: Measuring the Performance of Instant Acknowledgments in QUIC Handshakes}}. In \bibinfo{booktitle}{\emph{Proc. of ACM Internet Measurement Conference (IMC)}}. \bibinfo{publisher}{ACM}, \bibinfo{address}{New York}, \bibinfo{pages}{389--400}.
\newblock
\urldef\tempurl%
\url{https://doi.org/10.1145/3646547.3689022}
\showURL{%
\tempurl}


\bibitem[Nawrocki et~al\mbox{.}(2021)]%
        {nhsw-qqqrs-21}
\bibfield{author}{\bibinfo{person}{Marcin Nawrocki}, \bibinfo{person}{Raphael Hiesgen}, \bibinfo{person}{Thomas~C. Schmidt}, {and} \bibinfo{person}{Matthias W{\"a}hlisch}.} \bibinfo{year}{2021}\natexlab{}.
\newblock \showarticletitle{{QUICsand: Quantifying QUIC Reconnaissance Scans and DoS Flooding Events}}. In \bibinfo{booktitle}{\emph{Proc. of ACM Internet Measurement Conference (IMC)}}. \bibinfo{publisher}{ACM}, \bibinfo{address}{New York}, \bibinfo{pages}{283--291}.
\newblock
\urldef\tempurl%
\url{https://doi.org/10.1145/3487552.3487840}
\showURL{%
\tempurl}


\bibitem[Nawrocki et~al\mbox{.}(2022)]%
        {nthms-ibtcq-22}
\bibfield{author}{\bibinfo{person}{Marcin Nawrocki}, \bibinfo{person}{Pouyan~Fotouhi Tehrani}, \bibinfo{person}{Raphael Hiesgen}, \bibinfo{person}{Jonas M{\"u}cke}, \bibinfo{person}{Thomas~C. Schmidt}, {and} \bibinfo{person}{Matthias W{\"a}hlisch}.} \bibinfo{year}{2022}\natexlab{}.
\newblock \showarticletitle{{On the Interplay between TLS Certificates and QUIC Performance}}. In \bibinfo{booktitle}{\emph{Proc. of 18th International Conference on emerging Networking EXperiments and Technologies (CoNEXT)}}. \bibinfo{publisher}{ACM}, \bibinfo{address}{New York, NY, USA}, \bibinfo{pages}{204--213}.
\newblock
\urldef\tempurl%
\url{https://dl.acm.org/doi/10.1145/3555050.3569123}
\showURL{%
\tempurl}


\bibitem[Piraux et~al\mbox{.}(2018)]%
        {pdb-oeoqi-18}
\bibfield{author}{\bibinfo{person}{Maxime Piraux}, \bibinfo{person}{Quentin {De Coninck}}, {and} \bibinfo{person}{Olivier Bonaventure}.} \bibinfo{year}{2018}\natexlab{}.
\newblock \showarticletitle{Observing the Evolution of QUIC Implementations}. In \bibinfo{booktitle}{\emph{Proc. of the Workshop on the Evolution, Performance, and Interoperability of QUIC}} (Heraklion, Greece) \emph{(\bibinfo{series}{EPIQ'18})}. \bibinfo{publisher}{ACM}, \bibinfo{address}{New York, NY, USA}, \bibinfo{pages}{8--14}.
\newblock
\urldef\tempurl%
\url{https://doi.org/10.1145/3284850.3284852}
\showDOI{\tempurl}


\bibitem[Richter et~al\mbox{.}(2016)]%
        {rwvab-maocn-16}
\bibfield{author}{\bibinfo{person}{Philipp Richter}, \bibinfo{person}{Florian Wohlfart}, \bibinfo{person}{Narseo Vallina-Rodriguez}, \bibinfo{person}{Mark Allman}, \bibinfo{person}{Randy Bush}, \bibinfo{person}{Anja Feldmann}, \bibinfo{person}{Christian Kreibich}, \bibinfo{person}{Nicholas Weaver}, {and} \bibinfo{person}{Vern Paxson}.} \bibinfo{year}{2016}\natexlab{}.
\newblock \showarticletitle{{A Multi-Perspective Analysis of Carrier-Grade NAT Deployment}}. In \bibinfo{booktitle}{\emph{Proc. of ACM IMC}}. \bibinfo{publisher}{ACM}, \bibinfo{address}{New York, NY, USA}, \bibinfo{pages}{215--229}.
\newblock
\urldef\tempurl%
\url{https://doi.org/10.1145/2987443.2987474}
\showURL{%
\tempurl}


\bibitem[R{\"u}th et~al\mbox{.}(2018)]%
        {rpdh-flaqi-18}
\bibfield{author}{\bibinfo{person}{Jan R{\"u}th}, \bibinfo{person}{Ingmar Poese}, \bibinfo{person}{Christoph Dietzel}, {and} \bibinfo{person}{Oliver Hohlfeld}.} \bibinfo{year}{2018}\natexlab{}.
\newblock \showarticletitle{{A First Look at QUIC in the Wild}}. In \bibinfo{booktitle}{\emph{Passive and Active Measurement}} \emph{(\bibinfo{series}{LNCS}, Vol.~\bibinfo{volume}{10771})}. \bibinfo{publisher}{Springer Nature}, \bibinfo{address}{Switzerland}, \bibinfo{pages}{255--268}.
\newblock


\bibitem[{Safaei Pour} et~al\mbox{.}(2023)]%
        {pnfb-csrim-23}
\bibfield{author}{\bibinfo{person}{Morteza {Safaei Pour}}, \bibinfo{person}{Christelle Nader}, \bibinfo{person}{Kurt Friday}, {and} \bibinfo{person}{Elias Bou-Harb}.} \bibinfo{year}{2023}\natexlab{}.
\newblock \showarticletitle{{A Comprehensive Survey of Recent Internet Measurement Techniques for Cyber Security}}.
\newblock \bibinfo{journal}{\emph{Computers \& Security}}  \bibinfo{volume}{128} (\bibinfo{year}{2023}), \bibinfo{pages}{103123}.
\newblock
\urldef\tempurl%
\url{https://doi.org/10.1016/j.cose.2023.103123}
\showDOI{\tempurl}


\bibitem[Schinazi and Rescorla(2023)]%
        {RFC-9368}
\bibfield{author}{\bibinfo{person}{D. Schinazi} {and} \bibinfo{person}{E. Rescorla}.} \bibinfo{year}{2023}\natexlab{}.
\newblock \bibinfo{booktitle}{\emph{{Compatible Version Negotiation for QUIC}}}.
\newblock \bibinfo{type}{RFC} 9368. \bibinfo{institution}{IETF}.
\newblock
\urldef\tempurl%
\url{https://doi.org/10.17487/RFC9368}
\showDOI{\tempurl}


\bibitem[Seemann and Iyengar(2020)]%
        {si-aqit-20}
\bibfield{author}{\bibinfo{person}{Marten Seemann} {and} \bibinfo{person}{Jana Iyengar}.} \bibinfo{year}{2020}\natexlab{}.
\newblock \showarticletitle{Automating {QUIC} Interoperability Testing}. In \bibinfo{booktitle}{\emph{Proceedings of the 2020 Workshop on the Evolution, Performance, and Interoperability of QUIC, EPIQ@SIGCOMM 2020, Virtual Event, USA, August 10-14, 2020}}, \bibfield{editor}{\bibinfo{person}{J{\"{o}}rg Ott} {and} \bibinfo{person}{Lars Eggert}} (Eds.). \bibinfo{publisher}{{ACM}}, \bibinfo{address}{New York, NY, USA}, \bibinfo{pages}{8--13}.
\newblock
\urldef\tempurl%
\url{https://doi.org/10.1145/3405796.3405826}
\showDOI{\tempurl}


\bibitem[Shbair et~al\mbox{.}(2016)]%
        {scfc-mftih-16}
\bibfield{author}{\bibinfo{person}{Wazen~M. Shbair}, \bibinfo{person}{Thibault Cholez}, \bibinfo{person}{Jerome Francois}, {and} \bibinfo{person}{Isabelle Chrisment}.} \bibinfo{year}{2016}\natexlab{}.
\newblock \showarticletitle{A multi-level framework to identify HTTPS services}. In \bibinfo{booktitle}{\emph{NOMS 2016 - 2016 IEEE/IFIP Network Operations and Management Symposium}}. \bibinfo{publisher}{IFIP}, \bibinfo{address}{Istanbul, Turkey}, \bibinfo{pages}{240--248}.
\newblock
\urldef\tempurl%
\url{https://doi.org/10.1109/NOMS.2016.7502818}
\showDOI{\tempurl}


\bibitem[Shreedhar et~al\mbox{.}(2021)]%
        {sppb-eqpow-21}
\bibfield{author}{\bibinfo{person}{Tanya Shreedhar}, \bibinfo{person}{Rohit Panda}, \bibinfo{person}{Sergey Podanev}, {and} \bibinfo{person}{Vaibhav Bajpai}.} \bibinfo{year}{2021}\natexlab{}.
\newblock \showarticletitle{{Evaluating QUIC Performance over Web, Cloud Storage and Video Workloads}}.
\newblock \bibinfo{journal}{\emph{IEEE Transactions on Network and Service Management}} (\bibinfo{year}{2021}), \bibinfo{numpages}{16}~pages.
\newblock
\urldef\tempurl%
\url{https://doi.org/10.1109/TNSM.2021.3134562}
\showDOI{\tempurl}


\bibitem[Sommese et~al\mbox{.}(2020)]%
        {sbajr-muatm-20}
\bibfield{author}{\bibinfo{person}{Raffaele Sommese}, \bibinfo{person}{Leandro Bertholdo}, \bibinfo{person}{Gautam Akiwate}, \bibinfo{person}{Mattijs Jonker}, \bibinfo{person}{Roland van Rijswijk-Deij}, \bibinfo{person}{Alberto Dainotti}, \bibinfo{person}{KC Claffy}, {and} \bibinfo{person}{Anna Sperotto}.} \bibinfo{year}{2020}\natexlab{}.
\newblock \showarticletitle{{MAnycast2: Using Anycast to Measure Anycast}}. In \bibinfo{booktitle}{\emph{Proc. of ACM IMC}}. \bibinfo{publisher}{ACM}, \bibinfo{address}{New York, NY, USA}, \bibinfo{pages}{456--463}.
\newblock
\urldef\tempurl%
\url{https://doi.org/10.1145/3419394.3423646}
\showURL{%
\tempurl}


\bibitem[Stock(2020)]%
        {s-halbw-20}
\bibfield{author}{\bibinfo{person}{Terin Stock}.} \bibinfo{year}{2020}\natexlab{}.
\newblock \bibinfo{title}{High Availability Load Balancers with Maglev}.
\newblock
\newblock
\urldef\tempurl%
\url{https://blog.cloudflare.com/high-availability-load-balancers-with-maglev/}
\showURL{%
\tempurl}


\bibitem[Thimmaraju and Scheuermann(2021)]%
        {ts-cmiyc-21}
\bibfield{author}{\bibinfo{person}{Kashyap Thimmaraju} {and} \bibinfo{person}{Bj{\"o}rn Scheuermann}.} \bibinfo{year}{2021}\natexlab{}.
\newblock \showarticletitle{{Count Me If You Can: Enumerating QUIC Servers Behind Load Balancers}}. In \bibinfo{booktitle}{\emph{Proc. of ECEASST NetSys}}. \bibinfo{publisher}{ECEASST}, \bibinfo{address}{Dortmund, Germany}, \bibinfo{numpages}{5}~pages.
\newblock
\urldef\tempurl%
\url{http://dx.doi.org/10.14279/tuj.eceasst.80.1172.1077}
\showURL{%
\tempurl}


\bibitem[Tong et~al\mbox{.}(2018)]%
        {ttsm-nqtcb-18}
\bibfield{author}{\bibinfo{person}{Van Tong}, \bibinfo{person}{Hai~Anh Tran}, \bibinfo{person}{Sami Souihi}, {and} \bibinfo{person}{Abdelhamid Mellouk}.} \bibinfo{year}{2018}\natexlab{}.
\newblock \showarticletitle{A Novel QUIC Traffic Classifier Based on Convolutional Neural Networks}. In \bibinfo{booktitle}{\emph{2018 IEEE Global Communications Conference (GLOBECOM)}}. \bibinfo{publisher}{{{IEEE}}}, \bibinfo{address}{Abu Dhabi, UAE}, \bibinfo{pages}{1--6}.
\newblock
\urldef\tempurl%
\url{https://doi.org/10.1109/GLOCOM.2018.8647128}
\showDOI{\tempurl}


\bibitem[Vaere et~al\mbox{.}(2018)]%
        {vbkt-tbses-18}
\bibfield{author}{\bibinfo{person}{Piet~De Vaere}, \bibinfo{person}{Tobias B{\"{u}}hler}, \bibinfo{person}{Mirja K{\"{u}}hlewind}, {and} \bibinfo{person}{Brian Trammell}.} \bibinfo{year}{2018}\natexlab{}.
\newblock \showarticletitle{{Three Bits Suffice: Explicit Support for Passive Measurement of Internet Latency in {QUIC} and {TCP}}}. In \bibinfo{booktitle}{\emph{Proc. of ACM IMC}}. \bibinfo{publisher}{{ACM}}, \bibinfo{address}{New York, NY, USA}, \bibinfo{pages}{22--28}.
\newblock
\urldef\tempurl%
\url{https://dl.acm.org/citation.cfm?id=3278535}
\showURL{%
\tempurl}


\bibitem[Vermeulen et~al\mbox{.}(2020)]%
        {vrbff-dcdoi-20}
\bibfield{author}{\bibinfo{person}{Kevin Vermeulen}, \bibinfo{person}{Justin~P. Rohrer}, \bibinfo{person}{Robert Beverly}, \bibinfo{person}{Olivier Fourmaux}, {and} \bibinfo{person}{Timur Friedman}.} \bibinfo{year}{2020}\natexlab{}.
\newblock \showarticletitle{{Diamond-Miner: Comprehensive Discovery of the Internet's Topology Diamonds}}. In \bibinfo{booktitle}{\emph{Proc. of USENIX NSDI}}. \bibinfo{publisher}{USENIX Association}, \bibinfo{address}{USA}, \bibinfo{pages}{479--494}.
\newblock


\bibitem[Wolsing et~al\mbox{.}(2019)]%
        {wrwh-ppowo-19}
\bibfield{author}{\bibinfo{person}{Konrad Wolsing}, \bibinfo{person}{Jan R{\"u}th}, \bibinfo{person}{Klaus Wehrle}, {and} \bibinfo{person}{Oliver Hohlfeld}.} \bibinfo{year}{2019}\natexlab{}.
\newblock \showarticletitle{{A Performance Perspective on Web Optimized Protocol Stacks: TCP+TLS+HTTP/2 vs. QUIC}}. In \bibinfo{booktitle}{\emph{Proc. of the Applied Networking Research Workshop}} \emph{(\bibinfo{series}{ANRW '19})}. \bibinfo{publisher}{ACM}, \bibinfo{address}{New York, NY, USA}, \bibinfo{pages}{1--7}.
\newblock
\urldef\tempurl%
\url{https://doi.org/10.1145/3340301.3341123}
\showDOI{\tempurl}


\bibitem[Xfinity(2021)]%
        {xfinity-aup-21}
\bibfield{author}{\bibinfo{person}{Xfinity}.} \bibinfo{year}{2021}\natexlab{}.
\newblock \bibinfo{title}{{Acceptable Use Policy for Xfinity Internet}}.
\newblock
\newblock
\urldef\tempurl%
\url{https://www.xfinity.com/corporate/customers/policies/highspeedinternetaup}
\showURL{%
\tempurl}
\newblock
\shownote{Last Access: May 2025}.


\bibitem[Yang et~al\mbox{.}(2021)]%
        {yfjr-dlazt-21}
\bibfield{author}{\bibinfo{person}{Lixuan Yang}, \bibinfo{person}{Alessandro Finamore}, \bibinfo{person}{Feng Jun}, {and} \bibinfo{person}{Dario Rossi}.} \bibinfo{year}{2021}\natexlab{}.
\newblock \showarticletitle{Deep Learning and Zero-Day Traffic Classification: Lessons Learned From a Commercial-Grade Dataset}.
\newblock \bibinfo{journal}{\emph{IEEE Transactions on Network and Service Management}} \bibinfo{volume}{18}, \bibinfo{number}{4} (\bibinfo{year}{2021}), \bibinfo{pages}{4103--4118}.
\newblock
\urldef\tempurl%
\url{https://doi.org/10.1109/TNSM.2021.3122940}
\showDOI{\tempurl}


\bibitem[Zirngibl et~al\mbox{.}(2021)]%
        {zbsja-io9ae-21}
\bibfield{author}{\bibinfo{person}{Johannes Zirngibl}, \bibinfo{person}{Philippe Buschmann}, \bibinfo{person}{Patrick Sattler}, \bibinfo{person}{Benedikt Jaeger}, \bibinfo{person}{Juliane Aulbach}, {and} \bibinfo{person}{Georg Carle}.} \bibinfo{year}{2021}\natexlab{}.
\newblock \showarticletitle{{It's over 9000: Analyzing Early QUIC Deployments with the Standardization on the Horizon}}. In \bibinfo{booktitle}{\emph{Proc. of ACM IMC}}. \bibinfo{publisher}{ACM}, \bibinfo{address}{New York, NY, USA}, \bibinfo{pages}{261–275}.
\newblock
\urldef\tempurl%
\url{https://doi.org/10.1145/3487552.3487826}
\showURL{%
\tempurl}


\bibitem[Zirngibl et~al\mbox{.}(2024)]%
        {zgssc-qhfqd-24}
\bibfield{author}{\bibinfo{person}{Johannes Zirngibl}, \bibinfo{person}{Florian Gebauer}, \bibinfo{person}{Patrick Sattler}, \bibinfo{person}{Markus Sosnowski}, {and} \bibinfo{person}{Georg Carle}.} \bibinfo{year}{2024}\natexlab{}.
\newblock \showarticletitle{{QUIC Hunter: Finding QUIC Deployments and Identifying Server Libraries Across the Internet}}. In \bibinfo{booktitle}{\emph{Passive and Active Measurement: 25th International Conference, PAM 2024, Virtual Event, March 11–13, 2024, Proceedings, Part II}}. \bibinfo{publisher}{Springer-Verlag}, \bibinfo{address}{Berlin, Heidelberg}, \bibinfo{pages}{273–290}.
\newblock
\showISBNx{978-3-031-56251-8}
\urldef\tempurl%
\url{https://doi.org/10.1007/978-3-031-56252-5_13}
\showURL{%
\tempurl}


\end{thebibliography}
